\documentclass[
  aps,
  pre,
  twocolumn,
  groupedaddress,
  floatfix,
  longbibliography,
  superscriptaddress,
  10pt
]{revtex4-2}

\usepackage{amsmath}
\usepackage{amssymb}
\usepackage{amsfonts}
\usepackage{mathrsfs}
\usepackage{bm}
\usepackage{bbold}
\usepackage{braket}
\usepackage{slashed}
\usepackage{cancel}
\usepackage{accents}

\usepackage{graphicx}
\usepackage{epsfig}
\usepackage{subfigure}
\usepackage{multirow}
\usepackage{float}

\usepackage{feynmf}

\usepackage{times}

\usepackage[dvipsnames,svgnames]{xcolor}

\usepackage{hyperref}
\hypersetup{
  colorlinks=true,
  linkcolor=Blue,
  citecolor=Blue,
  urlcolor=Blue
}

\usepackage[normalem]{ulem}
\usepackage{simpler-wick}
\usepackage{orcidlink}

\DeclareUnicodeCharacter{039B}{\ensuremath{\Lambda}}

\newcommand{\bea}{\begin{eqnarray}}
\newcommand{\eea}{\end{eqnarray}}

\newcommand{\nn}{\nonumber}

\begin{document}

\title{A Game-Theoretic Framework for Resource Allocation under Institutional Asymmetry}

\author{Jorge David Casta\~no-Yepes~\orcidlink{0000-0002-8654-1304}}
\email{jorge.yepes@correounivalle.edu.co}
\affiliation{ Departamento de Física, Universidad del Valle, Ciudad Universitaria Meléndez,
Santiago de Cali 760032, Colombia}
\author{Marcela Herrera~\orcidlink{0000-0001-7894-9206}}
\email{alba.herrera@correounivalle.edu.co }
\affiliation{ Departamento de Física, Universidad del Valle, Ciudad Universitaria Meléndez,
Santiago de Cali 760032, Colombia}
\affiliation{Centre for Bioinformatics and Photonics—CIBioFi}
\author{Javier Parra-Domínguez~\orcidlink{0000-0002-1088-9152}}
\affiliation{BISITE Research Group, University of Salamanca, 37007, Salamanca, Spain}

\begin{abstract}
We develop a two-player knapsack game to investigate resource allocation under individual capacity constraints, competition over shared resources, and institutional asymmetry. Players select feasible subsets of resources, while their payoffs depend on competitive losses, allocation rules, and asymmetric institutional benefits and costs. We derive the aggregate welfare function and show that allocation coefficients redistribute payoffs without directly affecting total welfare, whereas institutional parameters influence welfare through their net contribution. Consequently, institutional arrangements with identical welfare effects may induce different strategic incentives and Nash equilibrium outcomes. We analyze how capacity constraints and institutional priority shape equilibrium configurations and their welfare properties. A central result concerns the distinction between the social efficiency and strategic stability of shared resources. We derive conditions under which sharing is socially preferable to exclusive allocation yet remains strategically unstable because the non-priority player can increase their payoff by abandoning the resource. Conversely, sharing may remain strategically stable even when exclusive allocation yields higher aggregate welfare. These findings demonstrate that social efficiency and strategic stability need not coincide under institutional asymmetry.
\end{abstract}

\maketitle

\section{ Introduction}

The allocation of scarce resources among agents who make decisions in a decentralised manner is one of the central problems of economics. When several agents compete for a common pool of resources under capacity constraints, each agent’s choices condition the opportunities and payoffs available to the others. The resulting allocation therefore reflects not only the individual decisions of the agents, but also the strategic interdependence among them and the institutional rules governing access to shared resources. {More broadly, recent research has explored complementary approaches to strategic decision-making, including quantum and evolutionary game theory, which extend the range of frameworks used to investigate complex interactions and decision-making processes~\cite{SanzMartin2025}.}

The knapsack problem is a classical problem in combinatorial optimization and computational complexity~\cite{Martello1990,Lucas2014}. In its standard formulation, a finite collection of objects is given, each characterized by a value and a weight, together with a knapsack of limited capacity. The problem consists of selecting a subset of objects whose total weight does not exceed the available capacity while maximizing their total value \cite{Martello1990}. Beyond its conventional optimization formulation, the knapsack problem has also been studied in strategic settings, including two-person and two-group knapsack games \cite{Wang2010TwoPerson,Wang2010TwoGroup}, as well as formulations that explicitly connect game theory with knapsack optimization \cite{Wong2015}. These developments establish a connection between constrained combinatorial optimization and strategic decision-making.

The relevance of the knapsack structure, however, extends beyond the search for an optimal subset. It can also be embedded in strategic settings in which multiple agents make resource-selection decisions subject to individual constraints. Existing work has examined several such formulations, including Stackelberg knapsack games, in which the order of decision-making plays a central role \cite{Pferschy2021}. More broadly, the knapsack problem illustrates how combinatorial optimization problems can be connected to strategic decision-making, where the choices of one agent may affect the opportunities and payoffs available to others.

Building on these game-theoretical formulations, we consider a setting in which players independently select feasible subsets of resources to maximise their individual payoffs. The capacity constraints define the admissible strategy spaces, while the overlap between the players' selections generates strategic interdependence. Unlike formulations that focus on the structure of the game or the order of strategic decisions, our approach places particular emphasis on how competition over shared resources and asymmetric institutional arrangements affect individual payoffs and aggregate welfare.

Within this framework, we introduce an interaction term that depends on the overlap between the resources selected by the players. A resource selected exclusively by one player contributes its full value to that player's payoff. When both players select the same resource, however, competition produces two distinct effects. The parameter $\lambda_k$ quantifies the fraction of the resource value lost through competition, while the coefficients $\alpha_k^i$ determine how the remaining value is distributed between the players. These mechanisms separate the dissipation of value from its redistribution: $\lambda_k$ determines the reduction in aggregate welfare associated with a shared resource, whereas $\alpha_k^i$ governs the division of the residual value, subject to $\alpha_k^a+\alpha_k^b=1$. This distinction provides a basis for examining how competitive intensity and payoff allocation affect strategic incentives and aggregate welfare independently.

From an economic standpoint, the model brings together three mechanisms that are analytically distinct. First, it incorporates a competitive loss: when two agents select the same resource, a fraction $\lambda_k$ of its value is dissipated. This effect can be interpreted in terms of rivalry and congestion in the use of shared resources, connecting the model with the literature on common-pool resources and club goods \cite{Hardin1968,Buchanan1965,Ostrom1990,CornesSandler1996}. Second, an allocation rule is introduced through the coefficients $\alpha_k^i$, which distribute the value remaining after competition. These coefficients can be interpreted as reflecting the agents' relative bargaining power, providing a conceptual connection with bargaining theory and cooperative game theory \cite{Nash1950,Shapley1953}. Third, institutional access asymmetry is captured by a benefit $\beta_k$ for the favoured agent and a cost $\gamma_k$ for the disadvantaged agent. This mechanism represents how differences in access rights, institutional arrangements, administrative or judicial decisions, and potentially illegitimate practices such as corruption or undue influence can affect the distribution of resources and payoffs~\cite{Coase1960,North1990,Williamson1985,Krueger1974,ShleiferVishny1993}.

These three mechanisms also admit a complementary interpretation in terms of externalities. The competitive loss $\lambda_k$ represents a negative externality associated with competition over shared resources, reducing aggregate surplus, whereas the aggregate welfare effect of institutional asymmetry depends on the difference between $\beta_k$ and $\gamma_k$~\cite{Pigou1920,Scitovsky1954,BuchananStubblebine1962}. This perspective helps clarify the distinction between individual incentives and aggregate welfare. The allocation coefficients $\alpha_k^i$ affect the distribution of payoffs without directly affecting total welfare, whereas the net contribution of the institutional parameters, $\beta_k-\gamma_k$, determines whether the asymmetric mechanism increases, preserves, or reduces aggregate welfare as defined in the model. The interaction between these mechanisms and the players' capacity constraints may lead to Nash equilibria that do not maximise aggregate welfare, illustrating the broader distinction between strategic equilibrium and welfare optimality~\cite{Maskin1999}.

A central feature of the framework is the distinction between the technological structure of the allocation problem, defined by the values and weights of the resources and the players' capacity constraints, and its strategic and institutional structure, determined by the rules governing competition, payoff allocation, and access. This distinction has direct implications for aggregate welfare: the allocation coefficients $\alpha_k^i$ affect individual payoffs but cancel from the welfare function, whereas the institutional parameters $\beta_k$ and $\gamma_k$ enter aggregate welfare only through their net difference, $\beta_k-\gamma_k$. Thus, institutional arrangements that yield the same aggregate welfare may nevertheless generate different individual incentives and equilibrium outcomes. The framework therefore separates the sources of welfare variation from those that affect how payoffs and strategic incentives are distributed among players.

The distinction between individual incentives and aggregate welfare provides the basis for analysing the welfare implications of strategic resource allocation under institutional asymmetry. We derive the aggregate welfare function and examine how competition, payoff allocation, and institutional access rules affect welfare as a function of the resources selected by the players. This analysis clarifies which mechanisms alter aggregate welfare directly and which primarily affect the distribution of payoffs and strategic incentives. It also provides a basis for examining whether welfare-improving resource allocations can arise as Nash equilibria under different institutional arrangements.

The manuscript is organized as follows. Section~\ref{sec:Theoretical Framework: A Knapsack Resource Allocation Game} introduces the theoretical framework, defining the two-player knapsack game and its payoff structure for exclusive resources, shared resources, and asymmetric access. It also establishes the expressions for total payoff and social welfare and defines the Nash equilibrium conditions. Section~\ref{sec:Game Characterization} characterizes the game through numerical resource-allocation scenarios, examining how competition, unequal payoff allocation, and asymmetric access affect payoff matrices and equilibrium outcomes. Section~\ref{sec:Institutional Asymmetry, Welfare, and Strategic Behavior} analyzes the conditions under which asymmetric access generates negative individual payoffs and distinguishes resource-level losses from their effects on total payoff and aggregate welfare. Section~\ref{sec:Social Optimum} investigates socially optimal allocations and examines when abandoning a shared resource is consistent with both social optimality and strategic stability. Finally, Section~\ref{sec:Summary and conclusions} summarizes the main findings and presents the conclusions.

\section{Theoretical Framework: A Knapsack Resource Allocation Game}\label{sec:Theoretical Framework: A Knapsack Resource Allocation Game}

The classical knapsack problem provides a natural mathematical description of resource allocation under a capacity constraint. Given a finite set of items, each characterized by a value and a weight, one seeks a feasible subset whose total weight does not exceed the available capacity and whose total value is maximized. In the present work, we use the same combinatorial structure for a different purpose. Rather than treating the knapsack as a single-agent optimization problem, we interpret its feasible configurations as the strategy space of several agents competing for a common set of resources. The resulting formulation is therefore a resource allocation game in which the capacity constraint determines the admissible decisions and the overlap between the players' selections generates the strategic interaction.

This game-theoretical interpretation of the knapsack problem is not new. In particular, Wang, Xing, and Fang introduced a two-person knapsack game in which two investors, each subject to an individual budget constraint, select projects from a common pool~\cite{Wang2010TwoPerson}. Each project has an associated cost and potential market profit. A project selected by only one investor yields its full profit to that investor, whereas the profit of a project selected by both investors is shared proportionally. The players choose portfolios according to best-response behavior and maximize individual objective functions that may depend linearly on the profits of both investors~\cite{Wang2010TwoPerson}. A related two-group knapsack game was subsequently formulated in which investors are divided into two groups, each investor has an individual budget, and members of the same group coordinate their choices so that a project is not selected by more than one investor within the group~\cite{Wang2010TwoGroup}. Projects selected by investors belonging to different groups may be shared according to the relative power of the groups. These models establish the knapsack structure as a natural setting for strategic resource allocation and provide conditions for the existence of pure Nash equilibria and, in simplified instances, an analysis of the price of anarchy~\cite{Wang2010TwoGroup}.

Other game-theoretical formulations have considered different strategic structures. Wong proposed a theoretical formulation connecting the knapsack problem with game theory through a bargaining and spatial interpretation of the feasible resource allocations~\cite{Wong2015}. More recently, Pferschy, Nicosia, Pacifici, and Schauer considered a Stackelberg version of the knapsack game in which the items are divided between a leader and a follower. The leader can modify the profits associated with its own items as incentives, after which the follower solves a knapsack problem over the complete item set. The resulting problem is formulated as a bilevel game and analyzed primarily from the perspective of computational complexity and algorithmic tractability~\cite{Pferschy2021}. These results show that strategic knapsack models can be constructed with substantially different interaction mechanisms, depending on whether the players compete for common projects, coordinate within groups, bargain over feasible allocations, or interact hierarchically. Our formulation adopts the common-resource interpretation of the two-person knapsack game, but uses it as a general framework for studying the strategic allocation of finite resources. 

\subsection{The Knapsack Game}\label{subsec:The Knapsack Game}

Let $\mathcal{R}=\{1,\ldots,n\}$ denote a finite set of resources. Each resource $k\in\mathcal{R}$ is characterized by a weight $w_k$ and a value $p_k$. The weight represents the amount of capacity required to allocate the resource, while the value represents the benefit associated with its allocation. We consider two players, denoted by $J_a$ and $J_b$, with capacities $C_a$ and $C_b$, respectively. A pure strategy of player $J_i$ is represented by a binary vector $\mathbf{x}^i=(x^i_1,\ldots,x^i_n)$, where $x^i_k=1$ if resource $k$ is selected by player $J_i$ and $x^i_k=0$ otherwise. The admissible strategies satisfy
\bea
\sum_{k=1}^{n}w_kx^i_k\leq C_i.
\eea

The feasible strategy space of player $J_i$ is consequently
\bea
\mathcal{S}_i=\left\{\mathbf{x}^i\in\{0,1\}^{n}\,\middle|\,\sum_{k=1}^{n}w_kx^i_k\leq C_i\right\}.
\eea
Each element of $\mathcal{S}_i$ represents a complete feasible allocation of the available resources by player $J_i$. A classical strategy profile is therefore an element of $\mathcal{S}_a\times\mathcal{S}_b$. The capacity constraints define the set of decisions that each player can make, while the common resource set establishes the possibility that two otherwise independent decisions may involve the same resource. In this sense, the knapsack constraint determines the structure of the individual strategy spaces, whereas the common set of resources establishes the possibility of strategic interaction between the players.

The game is therefore defined by the players, their feasible strategy spaces, and the common set of resources from which they independently construct their allocations. At this stage, the payoff associated with a selected resource has not yet been specified. In particular, whether the value of a resource is assigned to one player, shared between the players, or affected by competition depends on the resulting strategy profile and will be introduced below.

\subsubsection{Exclusive Resources}\label{subsubsec:Exclusive Resources}

For a given strategy profile $(\mathbf{x}^a,\mathbf{x}^b)$, the resource set can be partitioned according to the selection pattern generated by the two players. A resource $k$ is said to be exclusive to player $J_i$ when it is selected by $J_i$ and not selected by the other player $J_j$, with $j\neq i$. For binary strategy variables, this condition is represented by the product $x^i_k(1-x^j_k)$.

The two factors impose the two requirements for exclusivity simultaneously: $x^i_k=1$ requires player $J_i$ to select the resource, whereas $(1-x^j_k)=1$ requires player $J_j$ not to select it. Thus, the product vanishes whenever either of these conditions is not satisfied. The four possible combinations of the binary variables are given in Table~\ref{tab:exclusive_indicator}.

\begin{table}[h!]
\centering
\begin{tabular}{c c c c c}
\hline
$x^i_k$ & $x^j_k$ & $1-x^j_k$ & $x^i_k(1-x^j_k)$ & Interpretation \\
\hline
$0$ & $0$ & $1$ & $0$ & Neither player selects resource $k$ \\
$0$ & $1$ & $0$ & $0$ & Only player $J_j$ selects resource $k$ \\
$1$ & $0$ & $1$ & $1$ & Only player $J_i$ selects resource $k$ \\
$1$ & $1$ & $0$ & $0$ & Both players select resource $k$ \\
\hline
\end{tabular}
\caption{Interpretation of the binary product $x^i_k(1-x^j_k)$ as an indicator of exclusive resource selection by player $J_i$.}
\label{tab:exclusive_indicator}
\end{table}

The binary structure also allows the number of resources selected exclusively by player $J_i$ to be written directly as
\bea
N_i^{\mathrm{ex}}=\sum_{k=1}^{n}x^i_k(1-x^j_k),\qquad j\neq i.
\eea
More generally, if each resource carries a value $p_k$, the same indicator can be used to weight the contribution of each exclusive resource. In this way, the resource-selection structure and the resource values remain separated: the binary variables determine which resources belong to the exclusive subset, while $p_k$ determines their contribution to the payoff.

We then define the payoff contribution from exclusive resources as
\bea
P_i^{\mathrm{ex}}(\mathbf{x}^i,\mathbf{x}^j)=\sum_{k=1}^{n}p_kx^i_k(1-x^j_k),\qquad j\neq i.
\label{eq:exclusive_payoff}
\eea

\subsubsection{Shared Resources}\label{subsubsec:Shared Resources}

A resource is shared when it is selected simultaneously by both players. For a given strategy profile, shared resources are therefore characterized by the overlap of the two selected resource sets. For binary strategy variables, this overlap is represented by the product $x^a_kx^b_k$.

Unlike the exclusive indicator, which requires one player to select the resource and the other not to select it, the shared indicator requires both selection variables to be equal to one. Its four possible values are summarized in Table~\ref{tab:shared_indicator}.

\begin{table}[h!]
\centering
\begin{tabular}{c c c c}
\hline
$x^a_k$ & $x^b_k$ & $x^a_kx^b_k$ & Interpretation \\
\hline
$0$ & $0$ & $0$ & Neither player selects resource $k$ \\
$0$ & $1$ & $0$ & Only player $J_b$ selects resource $k$ \\
$1$ & $0$ & $0$ & Only player $J_a$ selects resource $k$ \\
$1$ & $1$ & $1$ & Both players select resource $k$ \\
\hline
\end{tabular}
\caption{Interpretation of the binary product $x^a_kx^b_k$ as an indicator of shared resource selection.}
\label{tab:shared_indicator}
\end{table}

The number of shared resources in a given strategy profile is consequently
\bea
N^{\mathrm{sh}}=\sum_{k=1}^{n}x^a_kx^b_k.
\eea
This quantity measures the overlap between the two resource selections. It also identifies the part of the strategy profile in which the players' decisions are directly coupled. The shared resources therefore differ from exclusive resources not through an additional strategic variable, but through the joint realization of the two binary decisions.

The value associated with a shared resource requires an additional specification because simultaneous selection introduces an interaction between the players. In the present formulation, this interaction is described through two distinct ingredients. The first accounts for a possible reduction of the total value generated by competition, while the second determines how the remaining value is allocated between the players.

We distinguish between the value lost through competition and the allocation of the value that remains. For each resource $k$, let $\lambda_k\in[0,1]$ denote the fraction of its value $p_k$ that is lost when both players select the resource. The value remaining after the competitive loss is therefore $(1-\lambda_k)p_k$. We then introduce allocation coefficients $\alpha_k^i$ to distribute this remaining value between the two players. Since the coefficients represent the fractions received by the two players, they satisfy
\bea
\alpha_k^a+\alpha_k^b=1,\qquad 0\leq\alpha_k^i\leq1.
\label{eq:allocation_constraint}
\eea
The value received by player $J_i$ from a shared resource $k$ is consequently
\bea
(1-\lambda_k)\alpha_k^ip_k.
\eea

The dependence on the strategy profile enters through the shared-resource indicator. We therefore define the shared-resource contribution to the payoff of player $J_i$ as
\bea
P_i^{\mathrm{sh}}(\mathbf{x}^i,\mathbf{x}^j)=\sum_{k=1}^{n}(1-\lambda_k)\alpha_k^ip_kx^a_kx^b_k.
\label{eq:shared_payoff}
\eea
%

\subsubsection{Asymmetric Access to Shared Resources}\label{subsubsec:Asymmetric Access to Shared Resources}

The allocation of a shared resource may depend not only on the competitive interaction between the players, but also on an asymmetry in the circumstances under which they access or select the resource. Such an asymmetry may arise from the temporal order of the decisions, priority of access, an earlier strategic commitment, access to relevant information at a particular stage of the game, a legally recognized right, a judicial decision, or a preferential position resulting from institutional or social arrangements. It may also arise from mechanisms that are not formally recognized as legitimate, including corruption, bribery, preferential treatment, or other forms of undue influence. The common feature of these situations is that, conditional on both players selecting the same resource, the two players may face different economic consequences depending on the mechanism that determines access to the resource.

To represent this asymmetry, we introduce a binary variable $\tau_k^i\in\{0,1\}$, where $\tau_k^i=1$ indicates that player $J_i$ obtains the asymmetric advantage associated with resource $k$. Since this mechanism concerns the interaction between players over a shared resource, it is activated only when both players select that resource. We therefore impose
\bea
\tau_k^a+\tau_k^b=x^a_kx^b_k.
\label{eq:asymmetry_constraint}
\eea
For $x^a_kx^b_k=0$, the constraint requires $\tau_k^a=\tau_k^b=0$, so that no asymmetric-access effect is generated when the resource is not shared. For $x^a_kx^b_k=1$, the constraint requires $\tau_k^a+\tau_k^b=1$, meaning that exactly one of the two players receives the asymmetric advantage. The variable $\tau_k^i$ therefore does not represent an additional choice concerning whether resource $k$ is selected. It identifies the player favored by the mechanism governing access to a resource that has already been selected by both players.

The economic effect associated with this asymmetry is distinct from the competitive loss and from the subsequent allocation of the remaining value. The parameter $\lambda_k$ determines the fraction of $p_k$ that is lost as a consequence of competition, leaving $(1-\lambda_k)p_k$ available for allocation. The coefficients $\alpha_k^i$ then determine how this remaining value is distributed between the two players. The asymmetric-access contribution introduced here does not modify either of these mechanisms. Instead, it represents an additional economic consequence generated by the circumstances that favor one player over the other. This additional effect may take the form of an economic benefit for the favored player, an economic cost for the non-favored player, or both.

We characterize this effect using two non-negative parameters, $\beta_k\geq0$ and $\gamma_k\geq0$. The parameter $\beta_k$ represents the additional benefit obtained by the player who receives the asymmetric advantage associated with resource $k$, whereas $\gamma_k$ represents the additional cost incurred by the player who does not receive that advantage. These parameters are introduced as general economic quantities and are not assumed to have any particular functional dependence on the resource value $p_k$. In a specific model, the asymmetric benefit or cost may depend on $p_k$, for example through a proportional relation, a nonlinear dependence, or another resource-dependent specification. Such a dependence is not imposed at the level of the general formulation. The corresponding contribution to the payoff of player $J_i$ is defined as
\bea
P_i^{\mathrm{asym}}(\mathbf{x}^i,\mathbf{x}^j)=\sum_{k=1}^{n}\left[\beta_k\tau_k^i-\gamma_k\tau_k^j\right],\qquad j\neq i.
\label{eq:asymmetry_payoff}
\eea
The first term, $\beta_k\tau_k^i$, contributes an additional benefit only to the player who obtains the asymmetric advantage, whereas the second term, $-\gamma_k\tau_k^j$, contributes an additional cost only to the player who does not obtain it. The two effects are associated with the same asymmetric mechanism but are not required to have the same magnitude or even to coexist. In particular, the asymmetric mechanism may generate a benefit without an explicit cost to the other player, a cost without an additional benefit to the favored player, or simultaneous benefits and costs for the two players.

The economic interpretation of $\beta_k$ and $\gamma_k$ depends on the mechanism responsible for the asymmetry. When $\beta_k>0$ and $\gamma_k=0$, the mechanism generates an additional benefit for the favored player without a direct additional cost for the other player. This case can represent preferential access, an earlier strategic commitment, privileged access to information, a legally recognized right, or a preferential position established by an institutional arrangement. The same parameterization can describe an advantage obtained through an illegitimate mechanism, such as corruption, bribery, or other forms of undue influence, provided that the mechanism generates an additional economic benefit for the favored player without an explicitly modeled cost for the other player.

When $\beta_k=0$ and $\gamma_k>0$, the mechanism does not provide an additional payoff to the favored player, but imposes an additional cost on the player who does not obtain the asymmetric advantage. This case is useful when the economically relevant consequence of the mechanism is a loss incurred by the non-favored player rather than a transfer of value to the favored player. The negative contribution should therefore be interpreted as an additional cost generated by the asymmetric mechanism, rather than as a negative value of the resource itself.

When both $\beta_k>0$ and $\gamma_k>0$, the asymmetric mechanism produces an additional gain for the favored player together with an additional cost for the non-favored player. This situation may arise when a judicial or institutional decision favors one party while imposing legal fees, compensation, sanctions, or other costs on the other party. It may also describe competitive or institutional processes in which the mechanism determining access produces consequences for both participants while simultaneously granting one of them a preferential position. The two parameters allow the magnitude of the additional gain and the additional cost to be specified independently.

The limiting case $\beta_k=\gamma_k=0$ removes the asymmetric-access contribution for resource $k$. The payoff associated with that resource is then determined only by its selection status, the competitive effect, and the allocation of the remaining value. More generally, allowing $\beta_k$ and $\gamma_k$ to depend on $k$ permits different resources to be associated with different institutional, legal, strategic, or economic mechanisms of asymmetry. A particular model may additionally impose a relation between these parameters and the resource characteristics, including its value $p_k$, but such a relation constitutes a model-specific assumption rather than a property of the general game.

It is worth noting that particular parameterizations of the asymmetric effects can lead to additional properties of the game. For instance, consider a model in which both the additional benefit and the additional cost are proportional to the value of the resource, with the same proportionality coefficient,
\bea
\beta_k=\gamma_k=\eta p_k,\qquad \eta\geq0.
\label{eq:zero_sum_asymmetry}
\eea
In this case, the asymmetric contribution associated with a shared resource represents a pure redistribution between the two players. Since $\tau_k^a+\tau_k^b=1$ whenever resource $k$ is shared, the asymmetric contributions satisfy
\bea
P_a^{\mathrm{asym}}+P_b^{\mathrm{asym}}=0.
\eea
Thus, the asymmetric component of the game is zero-sum: any additional benefit obtained by the player favored by the asymmetric mechanism is accompanied by an equal loss for the other player. The parameter $\eta$ determines the magnitude of this effect relative to the resource value $p_k$. Other choices of their dependence on the resource characteristics may instead generate a non-zero-sum asymmetric contribution.

\subsection{Total Payoff and Social Welfare}\label{subsec:Total Payoff and Social Welfare}

The total payoff of player $J_i$ is obtained by adding the three contributions introduced above. For $j\neq i$,
\bea
P_i(\mathbf{x}^i,\mathbf{x}^j)=P_i^{\mathrm{ex}}(\mathbf{x}^i,\mathbf{x}^j)+P_i^{\mathrm{sh}}(\mathbf{x}^i,\mathbf{x}^j)+P_i^{\mathrm{asym}}(\mathbf{x}^i,\mathbf{x}^j).
\label{eq:general_payoff}
\eea

Using Eqs.~\eqref{eq:exclusive_payoff}, \eqref{eq:shared_payoff}, and \eqref{eq:asymmetry_payoff}, this expression becomes
\bea
P_i(\mathbf{x}^i,\mathbf{x}^j)&=&\sum_{k=1}^{n}p_kx^i_k(1-x^j_k)+\sum_{k=1}^{n}(1-\lambda_k)\alpha_k^ip_kx^a_kx^b_k\nn\\
&+&\sum_{k=1}^{n}\left[\beta_k\tau_k^i-\gamma_k\tau_k^j\right],
\qquad j\neq i.
\label{eq:general_payoff_explicit}
\eea

The social welfare associated with a strategy profile is given by the sum of the individual payoffs,
\bea
W(\mathbf{x}^a,\mathbf{x}^b)=P_a(\mathbf{x}^a,\mathbf{x}^b)+P_b(\mathbf{x}^a,\mathbf{x}^b).
\label{eq:social_welfare}
\eea

Substituting Eq.~\eqref{eq:general_payoff_explicit} for both players yields
\bea
W(\mathbf{x}^a,\mathbf{x}^b)
&=&
\sum_{k=1}^{n}p_k\left[x^a_k(1-x^b_k)+x^b_k(1-x^a_k)\right]\nn\\
&+&\sum_{k=1}^{n}(1-\lambda_k)(\alpha_k^a+\alpha_k^b)p_kx^a_kx^b_k\nn\\
&+&\sum_{k=1}^{n}\left[(\beta_k-\gamma_k)(\tau_k^a+\tau_k^b)\right].
\label{eq:welfare_intermediate}
\eea

Applying the constraints $\alpha_k^a+\alpha_k^b=1$ and $\tau_k^a+\tau_k^b=x^a_kx^b_k$, we obtain
\bea
&&W(\mathbf{x}^a,\mathbf{x}^b)\nn\\
&=&
\sum_{k=1}^{n}
\Big\{
p_k\left[x^a_k+x^b_k-(1+\lambda_k)x^a_kx^b_k\right]
+(\beta_k-\gamma_k)x^a_kx^b_k
\Big\}.\nn\\
\label{eq:total_welfare}
\eea

A useful limiting case is obtained when the asymmetric terms satisfy Eq.~\eqref{eq:zero_sum_asymmetry}. Under this parametrization, the asymmetric mechanism is payoff-relevant at the individual level but welfare-neutral at the aggregate level. The gain accruing to the advantaged player is exactly matched by the loss borne by the other player, so that the mechanism operates as a redistribution of the surplus associated with the shared resource rather than as a source of additional surplus or destruction of it. Consequently, the incidence of the asymmetry affects the distribution of welfare across players without altering its aggregate level.

This property is not shared by the competitive interaction. In the general case described by Eq.~\eqref{eq:total_welfare}, joint selection entails a loss of $\lambda_k p_k$ from the value generated by resource $k$, while the asymmetric mechanism contributes the net term $\beta_k-\gamma_k$. The welfare effect of a shared resource is therefore governed by the balance between these two mechanisms. Competition constitutes a genuine destruction of surplus, whereas asymmetry can be either welfare-neutral, welfare-enhancing, or welfare-reducing, depending on the relative magnitudes of $\beta_k$ and $\gamma_k$. In particular, whenever the net effect of the asymmetric mechanism is insufficient to compensate for the competitive loss, the interaction generates a reduction in aggregate welfare.

\subsection{Nash Equilibrium}
\label{subsec:Nash Equilibrium}

The payoff functions defined in Eq.~\eqref{eq:general_payoff} determine the individual incentives of the players, whereas Eq.~\eqref{eq:total_welfare} characterizes the aggregate welfare associated with a strategy profile. These two criteria generally need not select the same outcome. A strategy profile may maximize aggregate welfare while remaining vulnerable to unilateral deviations, or may be strategically stable despite failing to maximize social welfare.

For a given strategy $\mathbf{x}^j$ of player $J_j$, the best-response correspondence of player $J_i$ is defined as the set of feasible strategies that maximize the payoff of player $J_i$ against $\mathbf{x}^j$. We denote this set by
\bea
\mathrm{BR}_i(\mathbf{x}^j)=\underset{\mathbf{x}^i\in\mathcal{S}_i}{\operatorname{arg\,max}}\,P_i(\mathbf{x}^i,\mathbf{x}^j).
\label{eq:best_response}
\eea
Here, $\operatorname{arg\,max}$ denotes the set of arguments, in this case feasible strategies $\mathbf{x}^i$, at which the maximum value of the payoff is attained. The set may contain more than one strategy when different feasible strategies yield the same maximum payoff. The indices satisfy $i,j\in\{a,b\}$ with $i\neq j$.

A strategy profile $(\mathbf{x}^{a*},\mathbf{x}^{b*})$ constitutes a Nash equilibrium if each player's strategy is a best response to the strategy of the other player,
\bea
\mathbf{x}^{a*}\in\mathrm{BR}_a(\mathbf{x}^{b*}),\qquad\mathbf{x}^{b*}\in\mathrm{BR}_b(\mathbf{x}^{a*}).
\label{eq:nash_equilibrium}
\eea

Equivalently, a profile is a Nash equilibrium if no player can increase their payoff through a unilateral feasible deviation. For every player $J_i$, with $J_j$ denoting the other player, let $\mathbf{x}^{i*}$ and $\mathbf{x}^{j*}$ denote the strategies of players $J_i$ and $J_j$ in the candidate equilibrium profile, respectively, while $\widetilde{\mathbf{x}}^i$ denotes an arbitrary alternative feasible strategy of player $J_i$. The Nash condition is then
\bea
P_i(\mathbf{x}^{i*},\mathbf{x}^{j*})\geq P_i(\widetilde{\mathbf{x}}^i,\mathbf{x}^{j*}).
\label{eq:nash_condition}
\eea
The inequality must hold for every $\widetilde{\mathbf{x}}^i\in\mathcal{S}_i$. For the present game, a unilateral deviation generally changes the interaction status of several resources simultaneously. A resource selected exclusively by the deviating player may become shared, a shared resource may become exclusive to the other player, and the corresponding asymmetric-access state may change as well. Consequently, the best-response problem cannot in general be reduced to the independent optimization of individual resource payoffs. The capacity constraint couples the selection decisions across resources, so that the best response is itself a constrained knapsack problem whose objective is determined by the strategy of the opponent.

For a candidate profile $(\mathbf{x}^{a*},\mathbf{x}^{b*})$, let the corresponding individual payoffs be
\bea
P_a^*=P_a(\mathbf{x}^{a*},\mathbf{x}^{b*}),\qquad P_b^*=P_b(\mathbf{x}^{a*},\mathbf{x}^{b*}).
\eea

\begin{table*}
\centering
\caption{Parameter values considered in the resource allocation scenarios. The scenarios are constructed sequentially by introducing competitive losses, unequal allocation, asymmetric benefits, and asymmetric costs.}
\label{tab:parameter_scenarios}
\begin{tabular}{cccccc}
\hline
Parameter
& Competition / Congestion
& Unequal Allocation
& Asymmetric Access
& Asymmetric Access with Cost
\\
\hline

$C_a,C_b$
& $7,7$
& $7,7$
& $7,7$
& $7,7$
\\

$n$
& $5$
& $5$
& $5$
& $5$
\\

$\mathbf{w}$
& $(1,2,2,3,4)$
& $(1,2,2,3,4)$
& $(1,2,2,3,4)$
& $(1,2,2,3,4)$
\\

$\mathbf{p}$
& $(3,2,1,5,7)$
& $(3,2,1,5,7)$
& $(3,2,1,5,7)$
& $(3,2,1,5,7)$
\\

$\boldsymbol{\lambda}$
& $(0.1,0.3,0.5,0.7,0.9)$
& $(0.1,0.3,0.5,0.7,0.9)$
& $(0.1,0.3,0.5,0.7,0.9)$
& $(0.1,0.3,0.5,0.7,0.9)$
\\

$\boldsymbol{\alpha}^a$
& $(0.5,0.5,0.5,0.5,0.5)$
& $(0.9,0.8,0.7,0.6,0.9)$
& $(0.9,0.8,0.7,0.6,0.9)$
& $(0.9,0.8,0.7,0.6,0.9)$
\\

$\boldsymbol{\alpha}^b$
& $(0.5,0.5,0.5,0.5,0.5)$
& $(0.1,0.2,0.3,0.4,0.1)$
& $(0.1,0.2,0.3,0.4,0.1)$
& $(0.1,0.2,0.3,0.4,0.1)$
\\

$\boldsymbol{\beta}$
& N/A
& N/A
& $(0.3,0.6,0.2,1,0.4)$
& $(0.3,0.6,0.2,1,0.4)$
\\

$\boldsymbol{\gamma}$
& N/A
& N/A
& $(0,0,0,0,0)$
& $(0.1,0.2,0.1,0.4,0.3)$
\\

Priority
& N/A
& N/A
& $(\text{a},\text{a},\text{a},\text{a},\text{a})$
& $(\text{a},\text{a},\text{a},\text{a},\text{a})$
\\

\hline
\end{tabular}
\end{table*}

\begin{figure*}
  \centering

  \subfigure[Competition / Congestion]
  {
    \includegraphics[scale=0.2304]{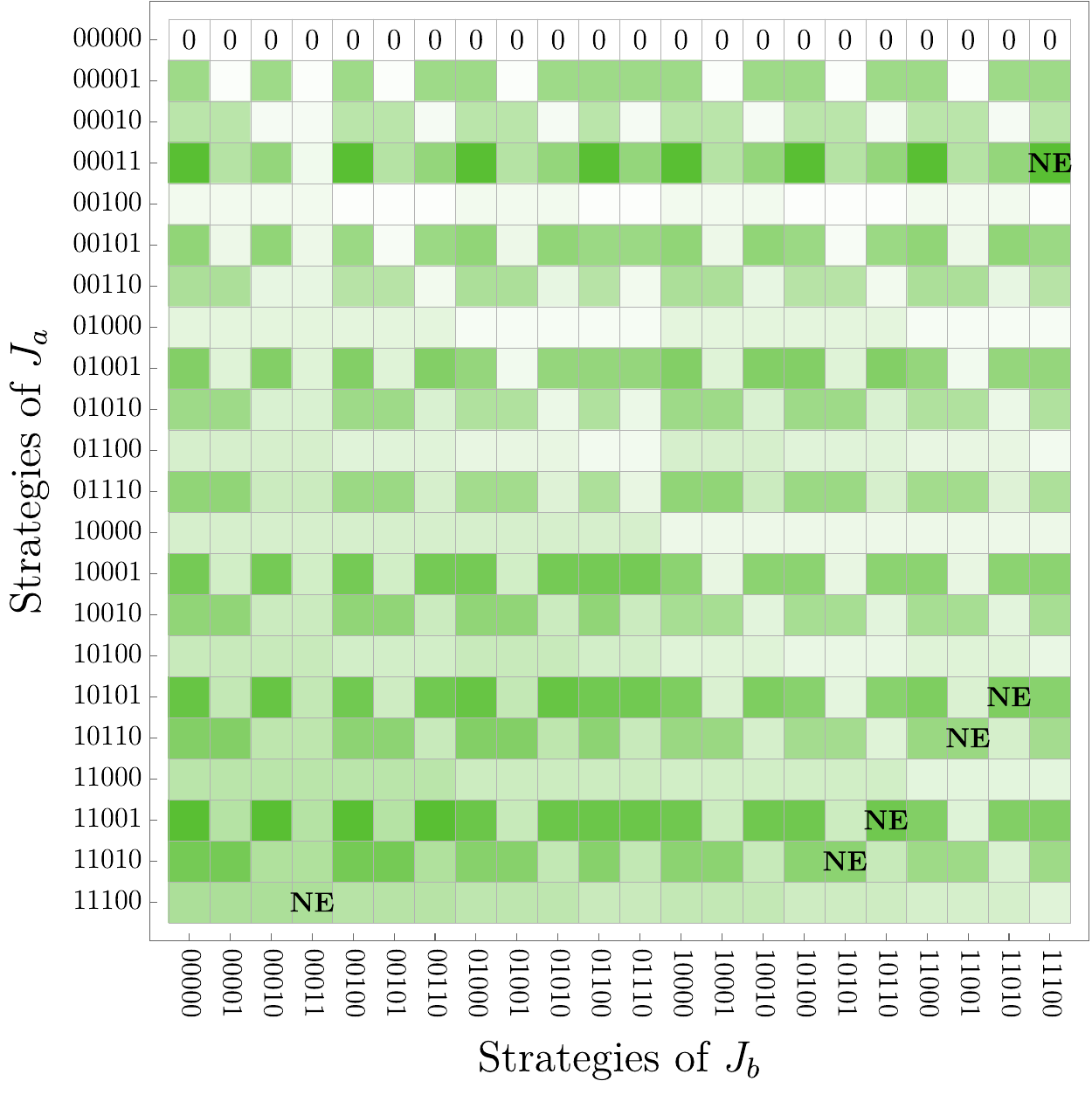}
    \label{fig:Pa-competition}
  }%
  \subfigure[Unequal Allocation]
  {
    \includegraphics[scale=0.204]{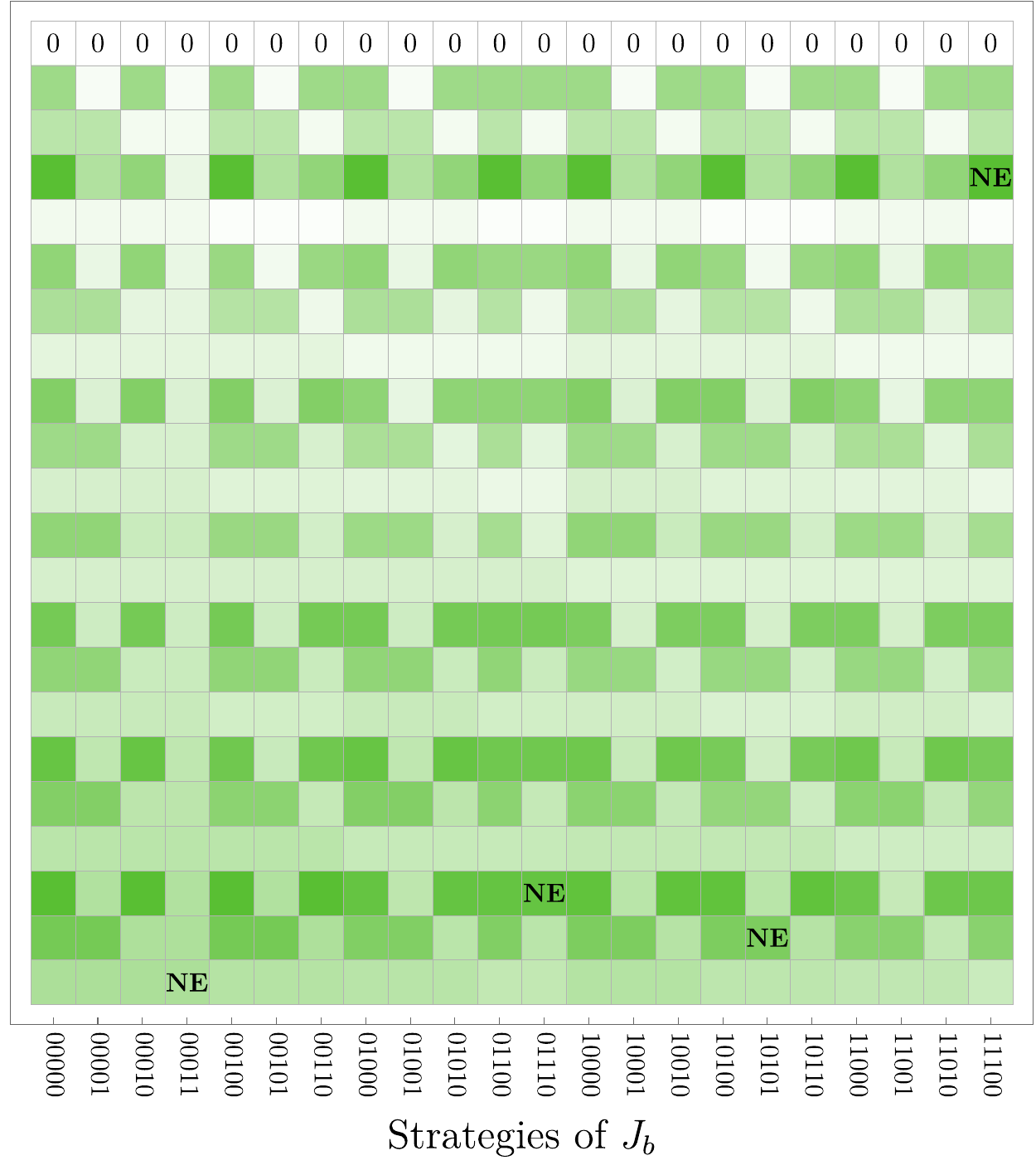}
    \label{fig:Pa-unequal}
  }%
  \subfigure[Asymmetric Access]
  {
    \includegraphics[scale=0.204]{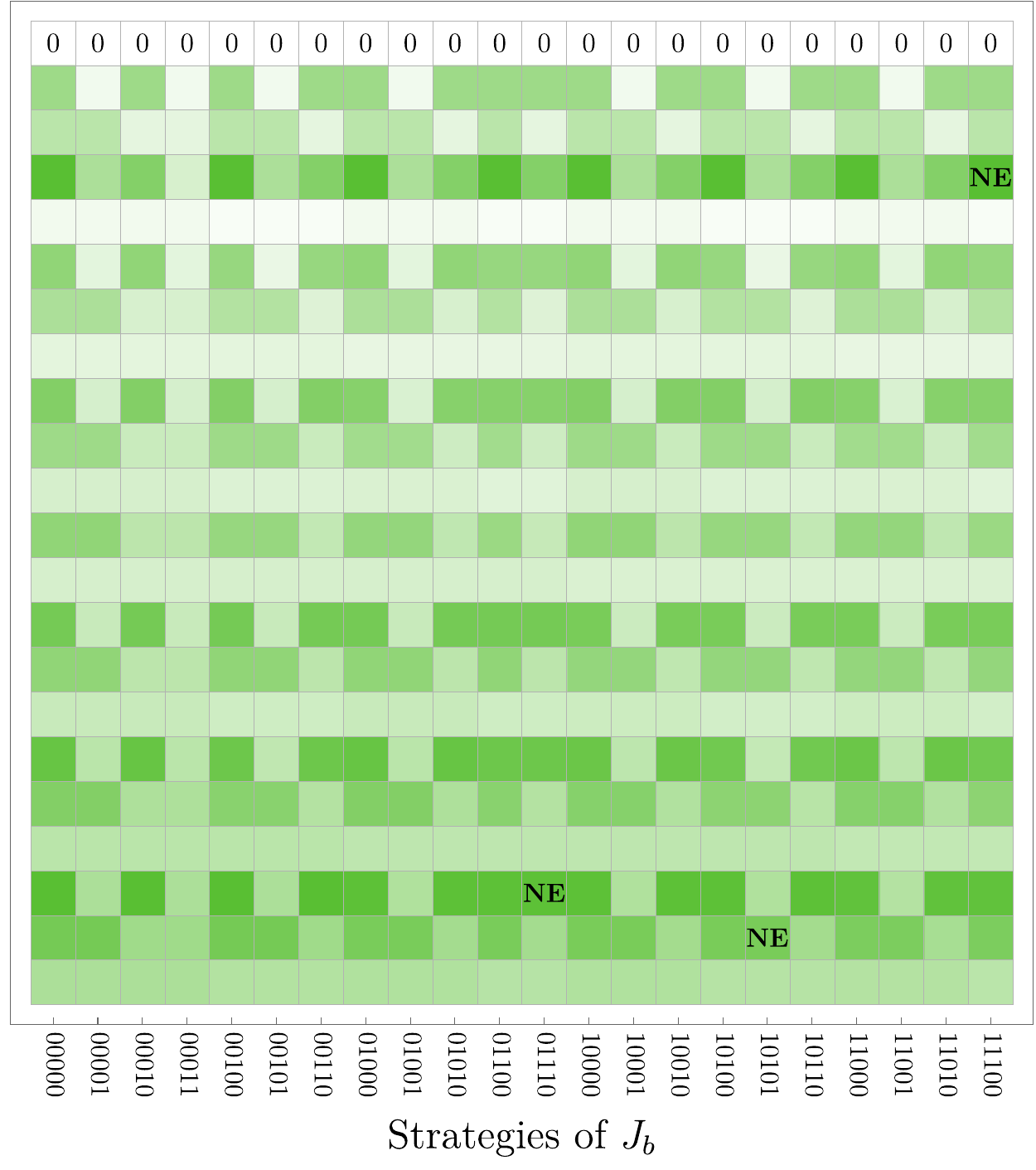}
    \label{fig:Pa-asymmetric}
  }%
  \subfigure[Asymmetric Access with Cost]
  {
    \includegraphics[scale=0.204]{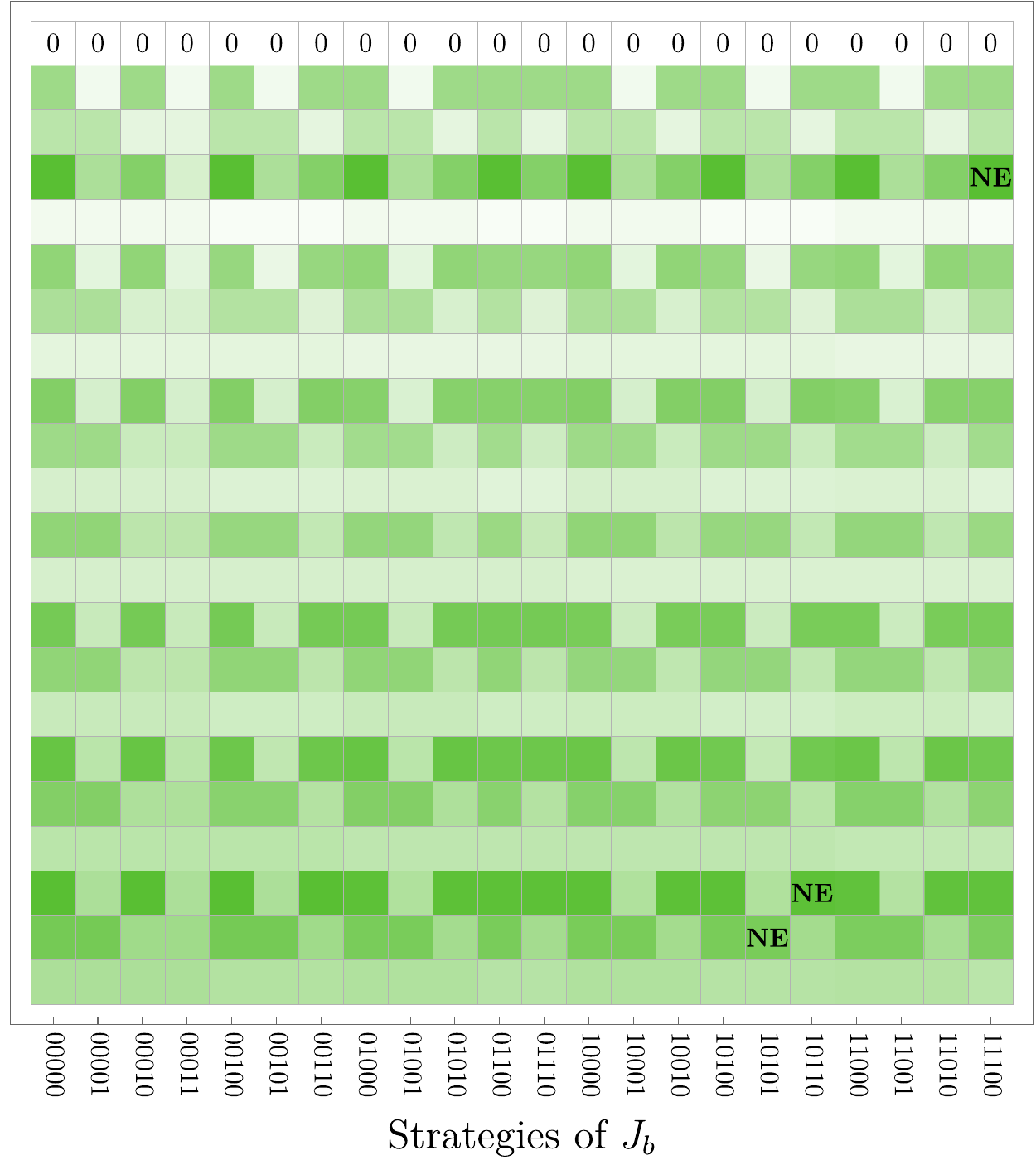}
    \label{fig:Pa-cost}
  }\\[0.2cm]

  \subfigure[Competition / Congestion]
  {
    \includegraphics[scale=0.2304]{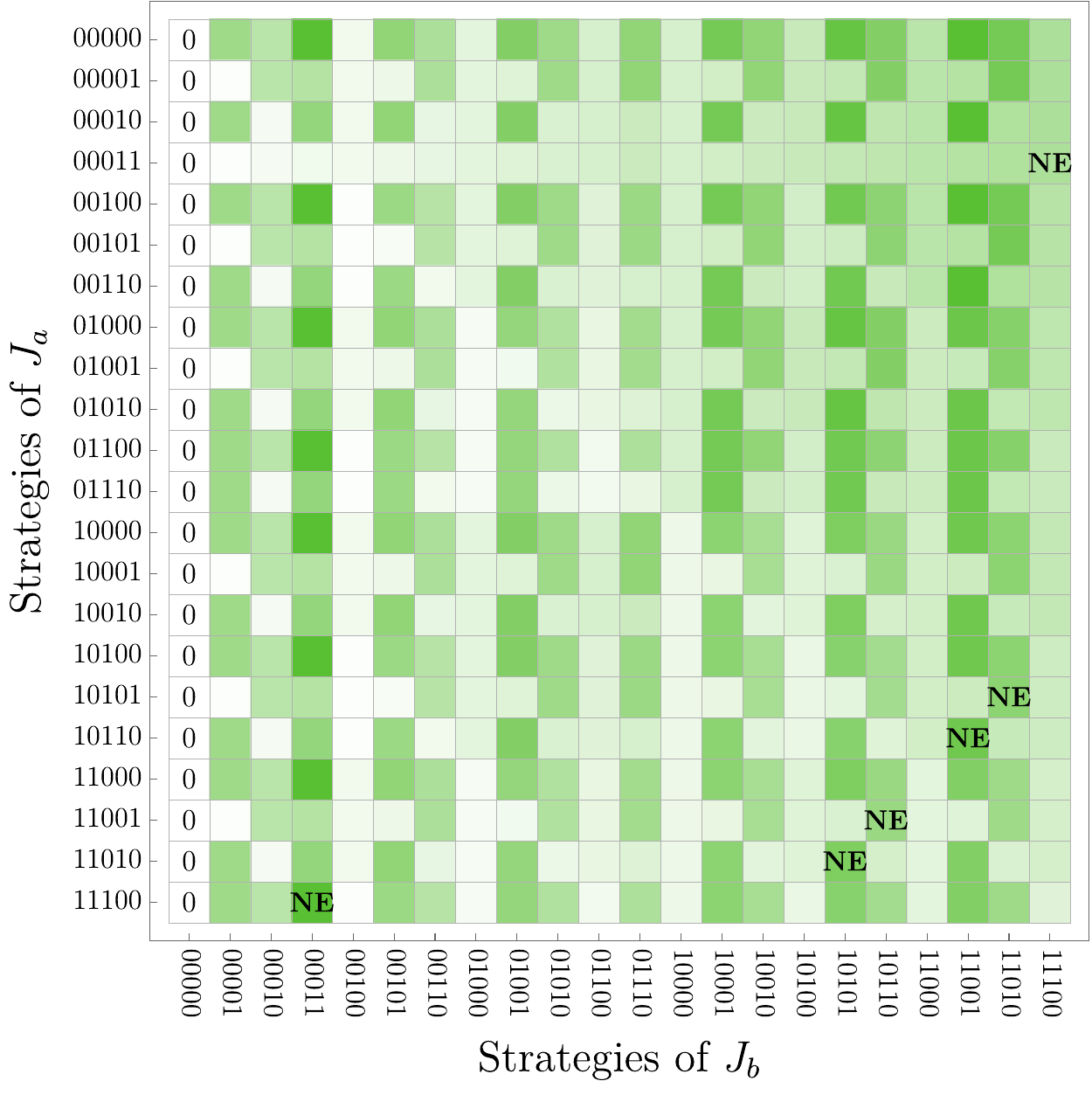}
    \label{fig:Pb-competition}
  }%
  \subfigure[Unequal Allocation]
  {
    \includegraphics[scale=0.204]{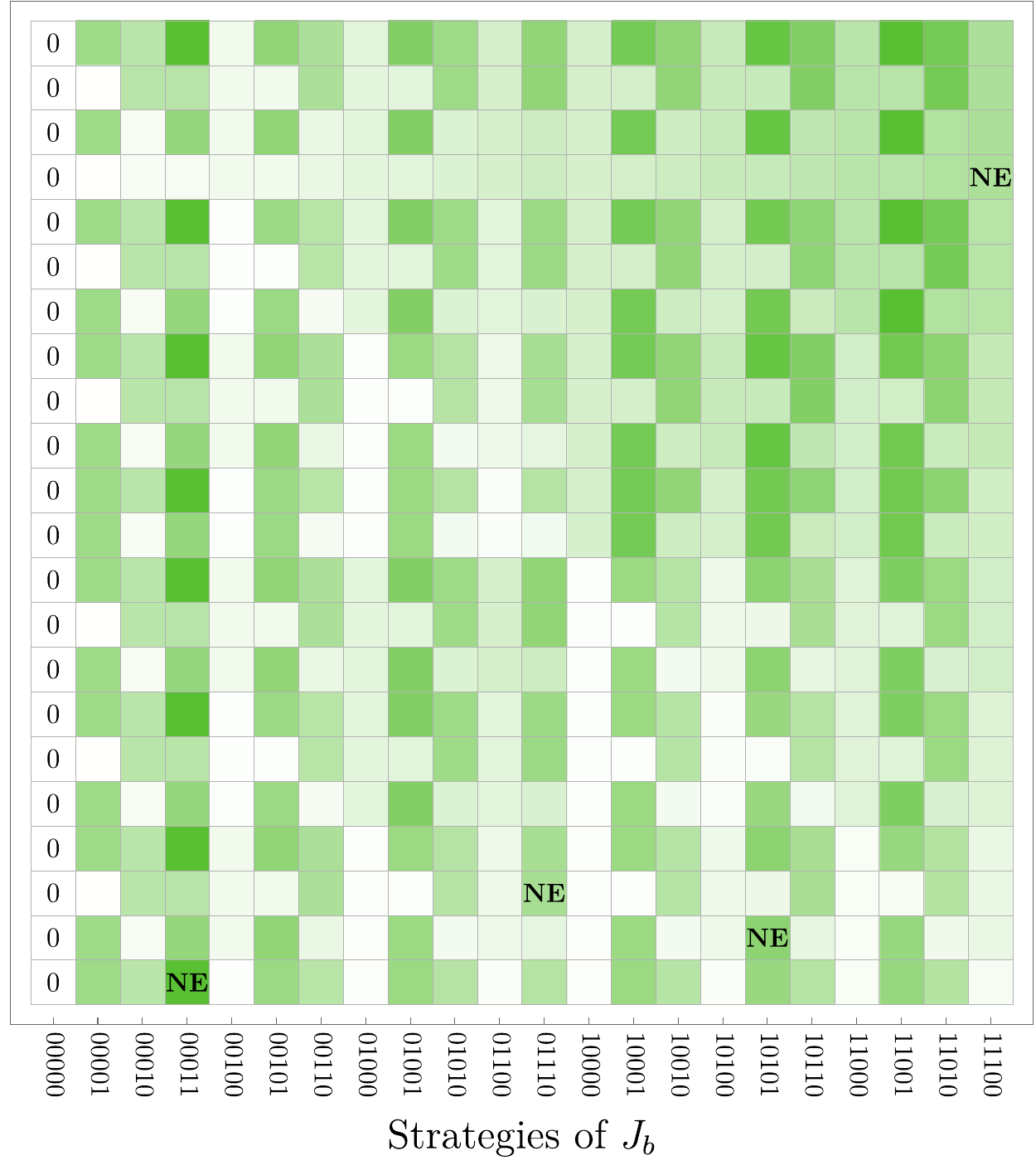}
    \label{fig:Pb-unequal}
  }%
  \subfigure[Asymmetric Access]
  {
    \includegraphics[scale=0.204]{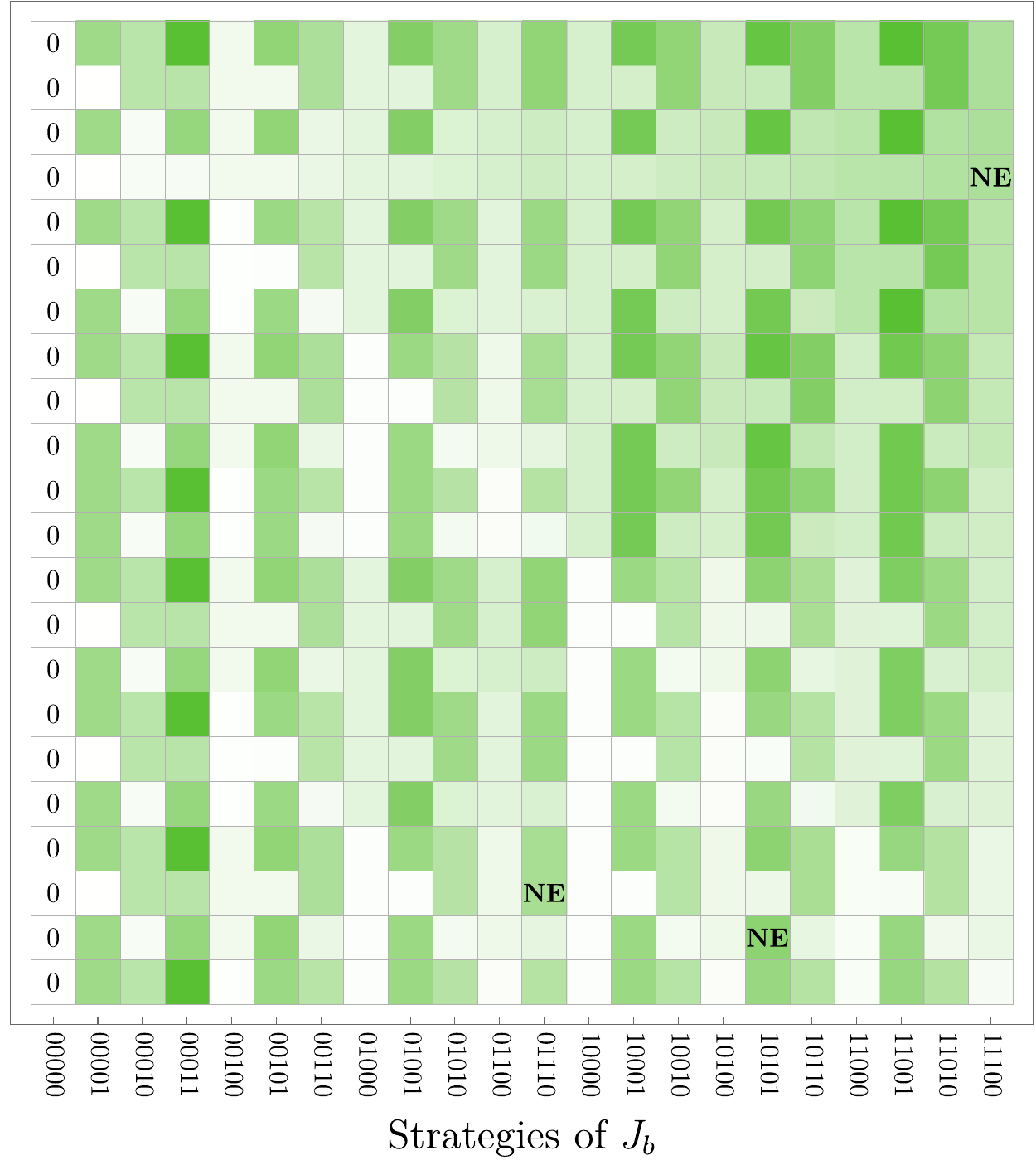}
    \label{fig:Pb-asymmetric}
  }%
  \subfigure[Asymmetric Access with Cost]
  {
    \includegraphics[scale=0.204]{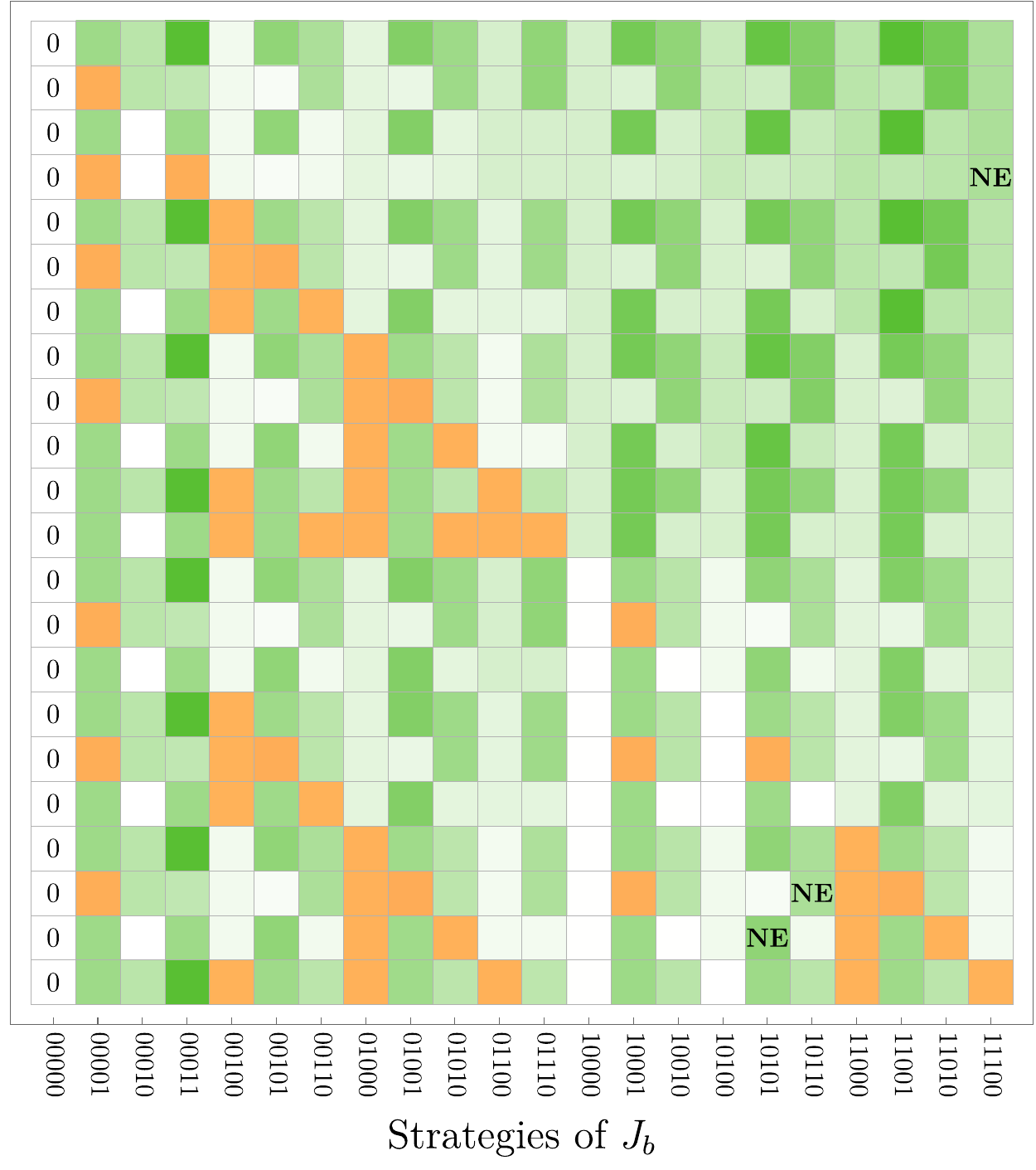}
    \label{fig:Pb-cost}
  }\\
\includegraphics[scale=0.35]{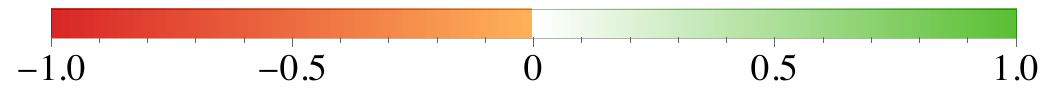}
  \caption{Normalized payoff matrices for the four resource allocation scenarios. The corresponding parameter values are given in Table~\ref{tab:parameter_scenarios}. The top and bottom rows show the normalized payoff matrices $P_a$ and $P_b$, respectively. The label $\textbf{NE}$ indicates pure-strategy Nash equilibria, while zeros in the cells denote payoffs that are exactly zero.}
  \label{fig:payoff_matrices}
\end{figure*}

\begin{table}
\centering
\caption{Pure-strategy Nash equilibria for the resource allocation scenarios defined in Table~\ref{tab:parameter_scenarios}. For each equilibrium, the corresponding strategies and payoffs of players $J_a$ and $J_b$ are reported.}
\label{tab:nash_equilibria}
\begin{tabular}{ccccc}
\hline
Scenario & Strategy A & Strategy B & $P_a$ & $P_b$ \\
\hline

\multirow{6}{*}{Competition}
&
$(0,0,0,1,1)$ & $(1,1,1,0,0)$ & $12.00$ & $6.00$ \\
&
$(1,0,1,0,1)$ & $(1,1,0,1,0)$ & $9.35$ & $8.35$ \\
&
$(1,0,1,1,0)$ & $(1,1,0,0,1)$ & $7.35$ & $10.35$ \\
&
$(1,1,0,0,1)$ & $(1,0,1,1,0)$ & $10.35$ & $7.35$ \\
&
$(1,1,0,1,0)$ & $(1,0,1,0,1)$ & $8.35$ & $9.35$ \\
&
$(1,1,1,0,0)$ & $(0,0,0,1,1)$ & $6.00$ & $12.00$ \\
\hline

\multirow{4}{*}{Unequal allocation}
&
$(0,0,0,1,1)$ & $(1,1,1,0,0)$ & $12.00$ & $6.00$ \\
&
$(1,1,0,0,1)$ & $(0,1,1,1,0)$ & $11.12$ & $6.28$ \\
&
$(1,1,0,1,0)$ & $(1,0,1,0,1)$ & $9.43$ & $8.27$ \\
&
$(1,1,1,0,0)$ & $(0,0,0,1,1)$ & $6.00$ & $12.00$ \\
\hline

\multirow{3}{*}{Asymmetric access}
&
$(0,0,0,1,1)$ & $(1,1,1,0,0)$ & $12.00$ & $6.00$ \\
&
$(1,1,0,0,1)$ & $(0,1,1,1,0)$ & $11.72$ & $6.28$ \\
&
$(1,1,0,1,0)$ & $(1,0,1,0,1)$ & $9.73$ & $8.27$ \\
\hline

\multirow{3}{*}{Access with cost}
&
$(0,0,0,1,1)$ & $(1,1,1,0,0)$ & $12.00$ & $6.00$ \\
&
$(1,1,0,0,1)$ & $(1,0,1,1,0)$ & $11.73$ & $6.17$ \\
&
$(1,1,0,1,0)$ & $(1,0,1,0,1)$ & $9.73$ & $8.17$ \\
\hline

\end{tabular}
\end{table}

The profitability of a unilateral deviation can then be characterized through the corresponding payoff differences. For player $J_a$, define
\bea
\Delta P_a(\widetilde{\mathbf{x}}^a;\mathbf{x}^{a*},\mathbf{x}^{b*})=P_a(\mathbf{x}^{a*},\mathbf{x}^{b*})-P_a(\widetilde{\mathbf{x}}^a,\mathbf{x}^{b*}).
\eea
Similarly, for player $J_b$,
\bea
\Delta P_b(\widetilde{\mathbf{x}}^b;\mathbf{x}^{a*},\mathbf{x}^{b*})=P_b(\mathbf{x}^{a*},\mathbf{x}^{b*})-P_b(\mathbf{x}^{a*},\widetilde{\mathbf{x}}^b).
\eea

The candidate profile is a Nash equilibrium if and only if neither player can obtain a higher payoff through any feasible unilateral deviation. Thus,
\bea
\Delta P_a(\widetilde{\mathbf{x}}^a;\mathbf{x}^{a*},\mathbf{x}^{b*})\geq0.
\label{eq:nash_payoff_difference_a}
\eea
This condition must hold for every $\widetilde{\mathbf{x}}^a\in\mathcal{S}_a$. Likewise,
\bea
\Delta P_b(\widetilde{\mathbf{x}}^b;\mathbf{x}^{a*},\mathbf{x}^{b*})\geq0.
\label{eq:nash_payoff_difference_b}
\eea
The latter condition must hold for every $\widetilde{\mathbf{x}}^b\in\mathcal{S}_b$. The structure of the payoff function makes the determinants of these deviations explicit. Changes in the allocation coefficients $\alpha_k^i$ alter the division of the value generated by shared resources and therefore affect the relative attractiveness of selecting resources already chosen by the opponent. The competitive parameters $\lambda_k$ modify the value associated with joint selection and can therefore discourage overlap when the corresponding loss is sufficiently large. Likewise, the asymmetric-access parameters $\beta_k$ and $\gamma_k$, together with the variables $\tau_k^i$, can alter the relative profitability of entering resources for which the opponent is already present. Hence, two parameter configurations may generate identical aggregate welfare for a given strategy profile while inducing different best responses and, consequently, different equilibrium sets.

This distinction is particularly relevant for asymmetric mechanisms satisfying Eq.~\eqref{eq:zero_sum_asymmetry}. In such a case, the asymmetric component is welfare-neutral at the aggregate level, but it remains payoff-relevant for each player. The redistribution generated by the mechanism can therefore modify strategic incentives and equilibrium selection without producing a direct change in social welfare. More generally, the equilibrium outcome depends on the individual incidence of the different payoff components, whereas welfare depends on their aggregate incidence as specified by Eq.~\eqref{eq:total_welfare}.

Since $\mathcal{S}_a$ and $\mathcal{S}_b$ are finite, the Nash conditions provide a finite characterization of the equilibrium set. In particular, for any fixed strategy of one player, the best-response problem of the other player can be evaluated over its feasible knapsack strategies. The intersections of the two best-response correspondences identify the pure-strategy Nash equilibria of the game.

\section{Game Characterization}\label{sec:Game Characterization}

Table~\ref{tab:parameter_scenarios} summarizes the parameter configurations considered in the four resource allocation scenarios, while Fig.~\ref{fig:payoff_matrices} shows the corresponding normalized payoff matrices for both players. Despite the successive modifications of the interaction rules, the figure shows a remarkably robust geometry of the payoff landscape. The regions of high and low payoff, as well as the overall organization of the strategy profiles, remain qualitatively similar across the four scenarios. The main changes are therefore localized modifications of the payoff values rather than a restructuring of the entire landscape. This indicates that the underlying combinatorial structure of the resource allocation problem remains visible in the payoff matrices even when the conditions governing competition and access are modified.

The Nash equilibria, however, are considerably more sensitive to these changes. As reported in Table~\ref{tab:nash_equilibria}, the number of pure-strategy Nash equilibria decreases from six in the Competition / Congestion scenario to four with Unequal Allocation and three with Asymmetric Access. The introduction of asymmetric costs does not change the number of equilibria, but it changes one of the equilibrium profiles. Thus, relatively modest modifications of the payoff landscape can produce substantial changes in the set of strategically stable allocations. This contrast between the robustness of the global payoff geometry and the sensitivity of the Nash equilibria reflects the local nature of equilibrium selection: a change that is small on the scale of the entire payoff matrix can nevertheless alter the ordering of payoffs among competing deviations and thereby eliminate or generate a Nash equilibrium.

The final scenario provides an additional qualitative change in the payoff structure. Negative payoff values appear in the matrix of $J_b$, the player with lower access to the shared resources, whereas the corresponding regions for $J_a$ remain nonnegative. These negative values indicate that, for some strategy profiles, the cost associated with the asymmetric access mechanism exceeds the payoff that $J_b$ can obtain from the corresponding allocation. The effect is therefore not merely a redistribution of positive payoffs between the players: under sufficiently unfavorable access conditions, the interaction can make some feasible allocations individually detrimental. The payoff matrices consequently distinguish between changes that modify how positive value is distributed and those that can make participation in a shared allocation economically unfavorable for the disadvantaged player.

More generally, the model shows that resource allocation is shaped by the interaction between scarcity and the institutional conditions governing access to resources. The underlying productive structure of the resources remains unchanged, yet changes in the rules of interaction can alter which feasible allocations are strategically sustainable and how the resulting value is distributed across players. This suggests that inefficiency need not originate from the resources themselves or from a lack of productive opportunities, but can emerge from the way agents compete over, share, or gain preferential access to them. In this sense, the model separates the technological structure of the allocation problem from its strategic and institutional structure, allowing the same set of resources to generate qualitatively different patterns of stable allocation as the conditions of access and interaction change.
\begin{table}
\centering
\caption{Parameter values for the asymmetric access with cost example with resource-dependent priority.}
\label{tab:asymmetric_cost_example}
\begin{tabular}{cc}
\hline
Parameter & Value \\
\hline
$C_a,C_b$ & $7,7$ \\
$n$ & $5$ \\
$\mathbf{w}$ & $(1,2,2,3,4)$ \\
$\mathbf{p}$ & $(3,2,1,5,7)$ \\
$\boldsymbol{\lambda}$ & $(0.1,0.3,0.5,0.7,0.9)$ \\
$\boldsymbol{\alpha}^a$ & $(0.9,0.8,0.7,0.6,0.9)$ \\
$\boldsymbol{\alpha}^b$ & $(0.1,0.2,0.3,0.4,0.1)$ \\
$\boldsymbol{\beta}$ & $(0.3,0.6,0.2,1.0,0.4)$ \\
$\boldsymbol{\gamma}$ & $(0.2,0.5,1.0,1.0,0.5)$ \\
Priority by resource & $(a,b,b,a,a)$ \\
\hline
\end{tabular}
\end{table}

\begin{figure}[h!]
    \centering
    \includegraphics[width=0.8\linewidth]{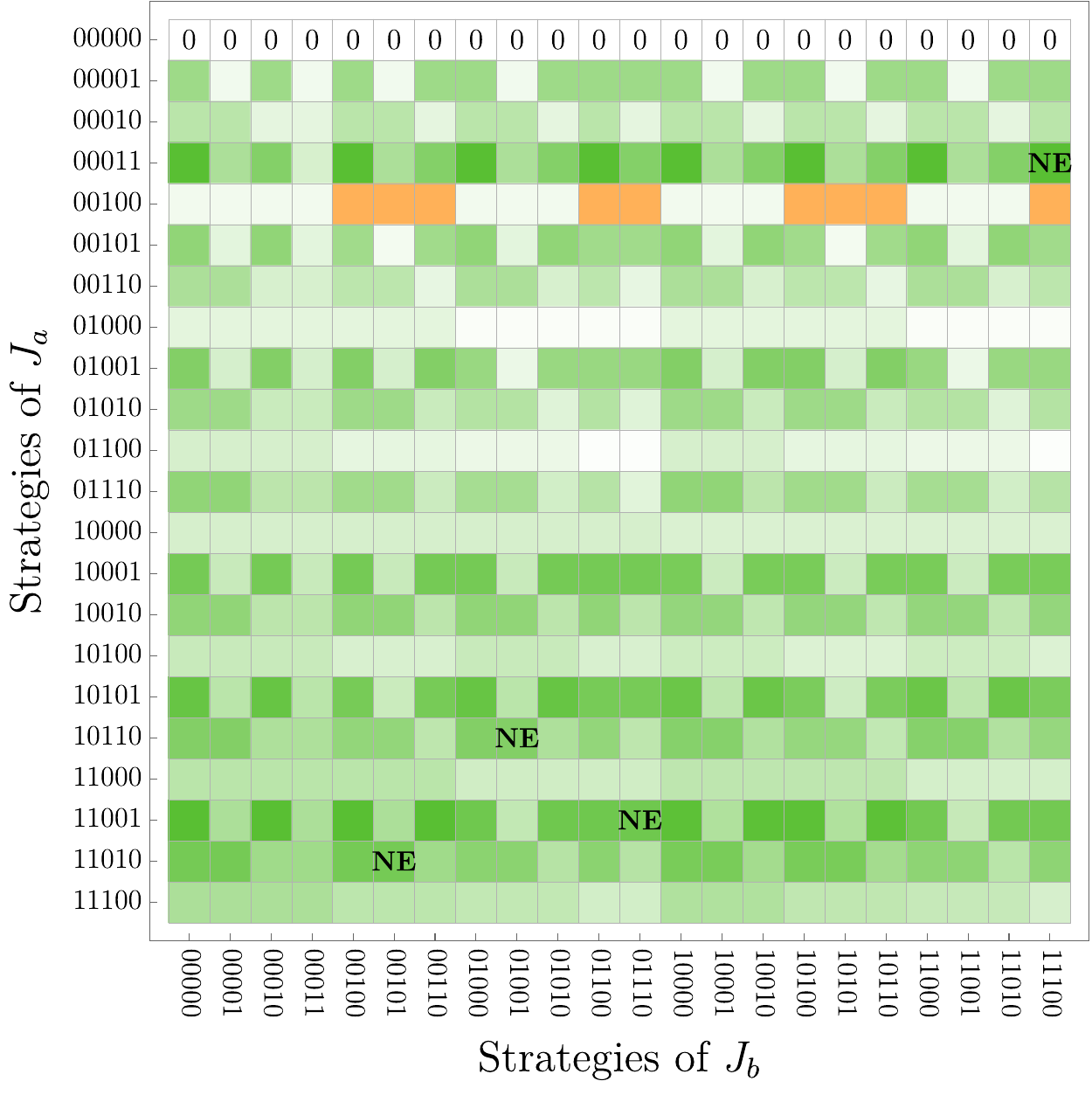}\\
    \vspace{0.3cm}
    \includegraphics[width=0.8\linewidth]{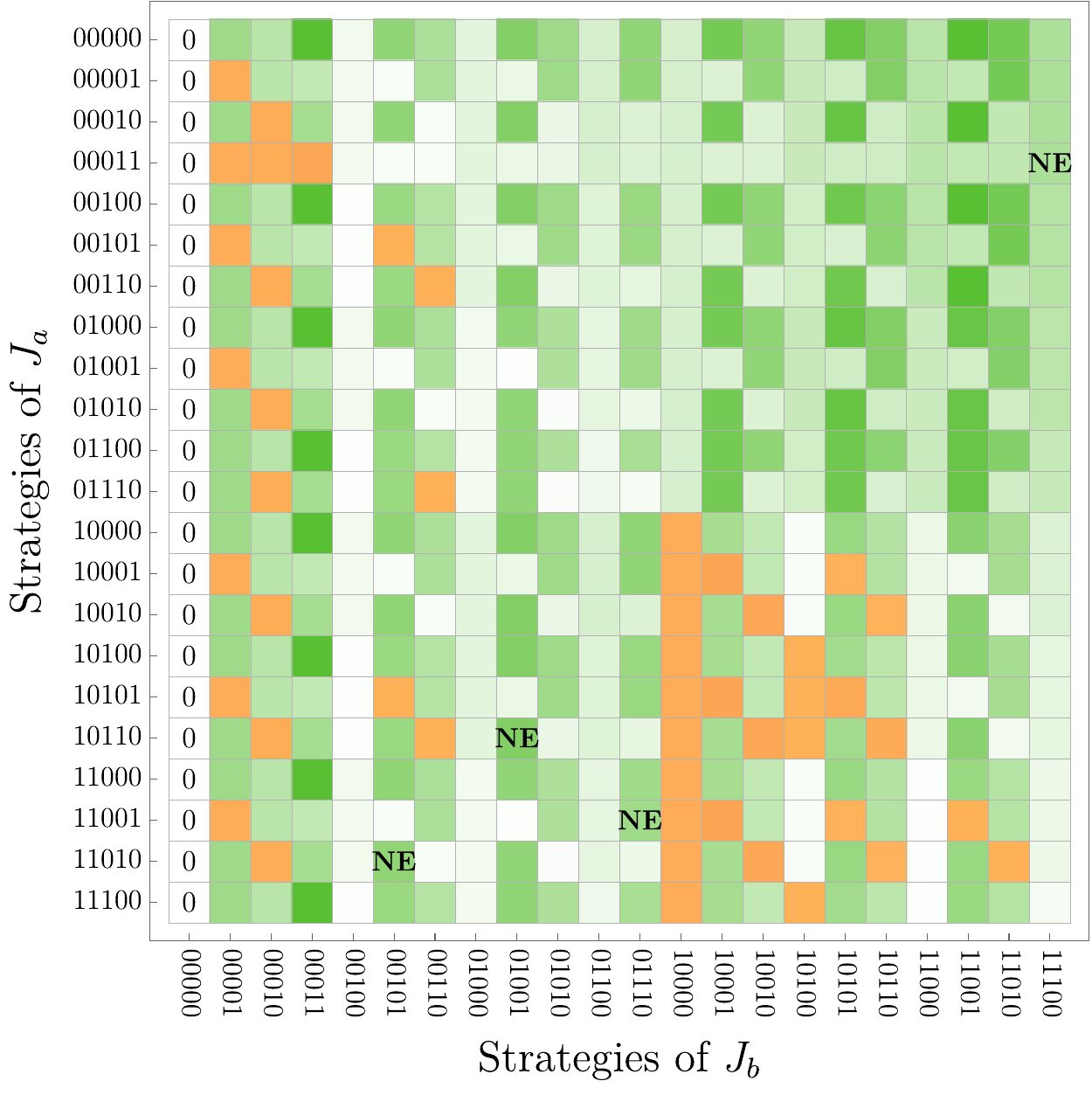}\\
    \vspace{0.3cm}
    \hspace{0.6cm}\includegraphics[scale=0.3]{Paper/Plots/commonLegend.pdf}
    \caption{Normalized payoff matrices for the asymmetric access with cost scenario. Parameters are given in Table~\ref{tab:asymmetric_cost_example}. The top and bottom panels show $P_a$ and $P_b$, respectively.}
    \label{fig:payoff_matrices-with-loss}
\end{figure}

In the previous example, priority was assigned exclusively to player $J_a$. Table~\ref{tab:asymmetric_cost_example} shows a scenario in which priority, additional benefits, and access costs vary across resources, allowing either player to occupy a privileged or non-priority position depending on the resource under consideration.

The payoff matrices show how resource-dependent priority modifies the strategic landscape of the game and creates a direct association between the players' resource-selection profiles and their total payoffs. Under the priority rule specified in Table~\ref{tab:asymmetric_cost_example}, $J_b$ has priority over resources $2$ and $3$, whereas $J_a$ has priority over resources $1$, $4$, and $5$. Thus, the profitability of a shared resource depends not only on its intrinsic value and competitive loss, but also on the player's position in the access hierarchy. For example, when both players select resource $3$, the contribution to the payoff of $J_a$ is $(1-\lambda_3)\alpha_3^a p_3-\gamma_3=-0.65$, whereas the corresponding contribution to $J_b$ is positive. Similarly, resources $2$, $4$, and $5$ can generate negative contributions for the corresponding non-priority player. The red regions therefore identify strategy profiles in which the losses associated with non-priority access are not compensated by the other resources selected by the player. From an economic perspective, this illustrates an allocation environment in which the profitability of a resource portfolio depends jointly on an agent's own selection, the selections of other agents, and heterogeneous access rights across resources. A resource may therefore be attractive in isolation but become unprofitable when its selection overlaps with that of another agent and places the player in a sufficiently costly non-priority position.

\section{ Institutional Asymmetry, Welfare, and Strategic Behavior}\label{sec:Institutional Asymmetry, Welfare, and Strategic Behavior}

The payoff structure allows the effects of asymmetric access to be characterized at both the resource and allocation levels. In particular, the mechanism can generate negative contributions to the payoff of a non-priority player, while its aggregate effect on social welfare depends on a different combination of parameters. This distinction provides a natural basis for separating the individual incidence of an institutional arrangement from its aggregate welfare effect and, subsequently, from its effect on strategic behavior and equilibrium.

The payoff structure allows the origin of negative resource-level payoffs to be characterized analytically. In particular, we ask {\it under what conditions can a shared resource generate a negative contribution to the payoff of one player while remaining non-negative, or even profitable, for the other?}

In order to answer the this question, consider a fixed resource $k$. The contribution of this resource to the payoff of player $J_i$ is
\bea
\pi_k^i=p_kx_k^i(1-x_k^j)+(1-\lambda_k)\alpha_k^ip_kx_k^ax_k^b+\beta_k\tau_k^i-\gamma_k\tau_k^j.\nn\\
\label{eq:resource_payoff}
\eea
Given the parameter restrictions $p_k\geq0$, $0\leq\lambda_k\leq1$, $0\leq\alpha_k^i\leq1$, $\beta_k\geq0$, and $\gamma_k\geq0$, all terms contributing to $\pi_k^i$ are nonnegative except for the access cost $-\gamma_k\tau_k^j$. Therefore, a negative contribution requires $\tau_k^j=1$. This condition can occur only when the resource is shared, implying $x_k^a=x_k^b=1$. The exclusive contribution then vanishes, while $\tau_k^i=0$ because the asymmetric advantage can be assigned to only one player. Consequently, the resource-level contribution to the payoff of the non-priority player becomes
\bea
\pi_k^i=(1-\lambda_k)\alpha_k^ip_k-\gamma_k.
\label{eq:nonpriority_resource_payoff}
\eea
Hence, the necessary and sufficient condition for resource $k$ to generate a negative contribution to the payoff of the non-priority player is
\bea
\pi_k^i<0\Longleftrightarrow\gamma_k>(1-\lambda_k)\alpha_k^ip_k.
\label{eq:general_resource_loss}
\eea
This condition identifies the threshold at which the cost associated with non-priority access exceeds the effective value obtained from the shared resource. The effective value is reduced relative to the original resource value $p_k$ by two mechanisms. First, competition removes the fraction $\lambda_k$ of the resource value. Second, the remaining value is divided according to $\alpha_k^i$. The resulting quantity $(1-\lambda_k)\alpha_k^ip_k$ is therefore the maximum contribution that the non-priority player receives from the shared resource before accounting for the asymmetric access cost.

The same condition can be written as a bound on any parameter entering the effective value. For $p_k>0$ and $0\leq\lambda_k<1$,
\bea
\alpha_k^i<\frac{\gamma_k}{(1-\lambda_k)p_k}.
\label{eq:alpha_loss_condition}
\eea
Similarly, for $\alpha_k^ip_k>0$, the condition can be expressed as
\bea
\lambda_k>1-\frac{\gamma_k}{\alpha_k^ip_k}.
\label{eq:lambda_loss_condition}
\eea
These equivalent forms show that a negative contribution can result from a sufficiently large access cost, a sufficiently strong competitive loss, a sufficiently small allocation coefficient, or a sufficiently small resource value. All of these mechanisms act by reducing the effective value available to the non-priority player relative to the cost imposed by unfavorable access.

The resource-level condition can also be interpreted in terms of a dimensionless measure of the severity of asymmetric access. Defining
\bea
\chi_k^i=\frac{\gamma_k}{(1-\lambda_k)\alpha_k^ip_k},
\label{eq:asymmetry_severity}
\eea
for $(1-\lambda_k)\alpha_k^ip_k>0$, the three relevant regimes are $\chi_k^i<1$, $\chi_k^i=1$, and $\chi_k^i>1$. In the first regime, the effective value received by the non-priority player exceeds the access cost and the resource contributes positively to the player's payoff. At $\chi_k^i=1$, the player is indifferent at the resource level. For $\chi_k^i>1$, the access cost exceeds the effective value and the resource generates a negative contribution. This representation separates the relative severity of the institutional cost from the absolute scale of the resource and is useful when comparing different resources with different values, competitive losses, or allocation coefficients.

Nevertheless, a negative contribution from an individual shared resource does not imply that the player's total payoff is negative. Let $J_i$ denote the priority player and $J_j$ the non-priority player. For a profile in which $J_i$ has priority over every shared resource, the payoff of $J_j$ can be written as
\bea
P_j=\sum_{k=1}^{n}p_kx_k^j(1-x_k^i)+\sum_{k=1}^{n}\left[(1-\lambda_k)\alpha_k^jp_k-\gamma_k\right]x_k^ax_k^b.\nn\\
\label{eq:nonpriority_total_payoff}
\eea
The first term is nonnegative and represents the value obtained from resources selected exclusively by $J_j$. The second term contains the contributions of shared resources and may contain both positive and negative terms. Therefore, the total payoff becomes negative only when the cumulative losses associated with shared resources exceed the value generated by the player's exclusive resources. Explicitly,
\bea
&&P_j<0\nn\\
&&\Longleftrightarrow\sum_{k=1}^{n}\left[\gamma_k-(1-\lambda_k)\alpha_k^jp_k\right]x_k^ax_k^b>\sum_{k=1}^{n}p_kx_k^j(1-x_k^i).\nn\\
\label{eq:total_loss_condition}
\eea

Equation~\eqref{eq:total_loss_condition} establishes the distinction between resource-level and aggregate negativity. A shared resource satisfying Eq.~\eqref{eq:general_resource_loss} produces a negative contribution to the non-priority player's payoff, but the total payoff remains nonnegative whenever the value obtained from exclusive resources is sufficient to compensate these losses. A negative total payoff requires instead that the aggregate cost of unfavorable access, net of the value appropriated from the corresponding shared resources, exceeds all positive contributions from resources that remain exclusive to the player.

This distinction is particularly relevant in the present knapsack environment because the resources cannot in general be evaluated independently. The capacity constraint couples the selection decisions, so that accepting one resource may prevent the selection of another. Consequently, the strategic relevance of a resource-level loss depends not only on its own contribution to the payoff but also on the opportunity cost associated with the complete feasible allocation. A resource with a negative individual contribution may nevertheless be retained if abandoning it does not permit the player to construct a more profitable feasible bundle. Conversely, a resource with a positive contribution may be excluded because the capacity required to select it could be used for a combination of resources with a higher total payoff.

The preceding analysis separates three distinct channels through which the interaction between players affects resource allocation. The competitive parameter $\lambda_k$ determines the fraction of the resource value that is lost when both players select the same resource, the coefficients $\alpha_k^i$ determine how the remaining value is distributed, and the asymmetric-access parameters $\beta_k$ and $\gamma_k$ determine additional benefits and costs associated with the mechanism governing access. While these components enter the individual payoff functions separately, their aggregate incidence is not necessarily equivalent to their strategic incidence.

To examine this distinction, recall the social welfare function
$W(\mathbf{x}^a,\mathbf{x}^b)$ defined in Eq.~\eqref{eq:social_welfare}.
Using the general payoff function and the restrictions
$\alpha_k^a+\alpha_k^b=1$ and $\tau_k^a+\tau_k^b=x_k^ax_k^b$, it can be
written as in Eq.~\eqref{eq:total_welfare}, which makes the different
aggregate roles of the three mechanisms explicit. For a resource selected exclusively by one player, the contribution to welfare is simply $p_k$. When the resource is selected jointly, the total contribution becomes $(1-\lambda_k)p_k+\beta_k-\gamma_k$. Thus, the allocation coefficients $\alpha_k^i$ do not affect aggregate welfare because they only determine how the residual value is divided between the two players. By contrast, $\lambda_k$ directly reduces the aggregate value generated by joint selection and therefore represents a genuine destruction of surplus.

The asymmetric-access parameters $\beta_k$ and $\gamma_k$ have a different aggregate interpretation. For a resource selected by both players, their contribution to welfare depends only on their difference. The corresponding marginal effects are
\bea
\frac{\partial W}{\partial\beta_k}=x_k^ax_k^b,\qquad\frac{\partial W}{\partial\gamma_k}=-x_k^ax_k^b.
\label{eq:welfare_asym_derivatives}
\eea
Thus, $\beta_k$ and $\gamma_k$ have opposite effects on aggregate welfare whenever the resource is jointly selected. An increase in $\beta_k$ increases the aggregate contribution of the shared resource, whereas an increase in $\gamma_k$ reduces it. The asymmetric mechanism can therefore have three qualitatively different aggregate effects. If $\beta_k>\gamma_k$, the mechanism increases the welfare contribution of a shared resource relative to the case without asymmetric terms. If $\beta_k=\gamma_k$, it is welfare-neutral. If $\beta_k<\gamma_k$, it reduces aggregate welfare.

This behavior differs fundamentally from that of the competitive parameter. Competition generates the loss $\lambda_kp_k$ whenever both players select the same resource, whereas asymmetric access contributes the net term $\beta_k-\gamma_k$. Hence, the welfare contribution of a shared resource is determined by the balance between the competitive destruction of surplus and the net effect of the institutional access rule. An institution can therefore partially compensate for the welfare loss generated by competition, leave it unchanged, or amplify it, depending on the relation between $\beta_k-\gamma_k$ and $\lambda_kp_k$.

The preceding result provides a natural distinction between welfare equivalence and strategic equivalence. Consider two institutional configurations characterized, for resource $k$, by $(\beta_k,\gamma_k)$ and $(\widetilde{\beta}_k,\widetilde{\gamma}_k)$. For a fixed strategy profile, the two configurations generate the same contribution to social welfare whenever
\bea
\beta_k-\gamma_k=\widetilde{\beta}_k-\widetilde{\gamma}_k.
\label{eq:welfare_equivalence}
\eea
Under this condition, the two configurations are welfare-equivalent with respect to the asymmetric mechanism. Importantly, Eq.~\eqref{eq:welfare_equivalence} does not require the individual parameters to coincide. Different combinations of benefits and costs can therefore generate the same aggregate welfare while producing different individual payoffs.

The welfare equivalence condition can be understood directly from the structure of Eq.~\eqref{eq:total_welfare}. The welfare function does not distinguish between an additional unit of benefit assigned to the priority player and an additional unit of cost imposed on the non-priority player except through their net difference. Consequently, an increase in $\beta_k$ can be compensated by an equal increase in $\gamma_k$ without changing the welfare associated with any strategy profile in which resource $k$ is shared. This property defines an entire family of institutional configurations that are observationally equivalent from the perspective of aggregate welfare. {
When the institution is welfare-neutral ($\beta_k = \gamma_k$), the rule redistributes incentives between players without altering the aggregate value of the shared resource, so that institutions that are observationally equivalent in efficiency may nevertheless sustain different allocations.
}

The individual payoff functions do not possess the same invariance. To see the strategic distinction, suppose that player $J_a$ is favored by the asymmetric mechanism for resource $k$. The asymmetric payoff received by player $J_a$ contains the term $\beta_k$, whereas the corresponding payoff of player $J_b$ contains the cost term $-\gamma_k$. Under two institutional configurations, the change in the asymmetric payoff of the favored player is therefore determined by $\beta_k-\widetilde{\beta}_k$, while the corresponding change for the non-priority player is determined by $-(\gamma_k-\widetilde{\gamma}_k)$. These changes can be nonzero even when their sum vanishes.

More generally, equality of the net institutional effect is not sufficient to guarantee equality of the individual payoff functions. A sufficient condition for strategic equivalence of the asymmetric mechanism is that the separate effects on the individual players coincide, namely
\bea
\beta_k=\widetilde{\beta}_k,\qquad\gamma_k=\widetilde{\gamma}_k.
\label{eq:strategic_equivalence}
\eea
Thus, the condition for welfare equivalence is weaker than the condition for strategic equivalence. Two institutional arrangements may therefore be indistinguishable in terms of aggregate welfare while inducing different private incentives. Since best responses depend on individual payoffs rather than on their sum, such configurations can generate different best-response correspondences and, consequently, different Nash equilibrium sets.

The distinction is not restricted to a comparison between two discrete institutional configurations. It can also be analyzed through a continuous institutional parameter. Let the asymmetric parameters be differentiable functions of a scalar parameter $\eta_k$,
\bea
\beta_k=\beta_k(\eta_k),\qquad\gamma_k=\gamma_k(\eta_k).
\label{eq:institutional_parameterization}
\eea
The response of aggregate welfare to a marginal change in the institutional parameter follows directly from Eq.~\eqref{eq:total_welfare},
\bea
\frac{\partial W}{\partial\eta_k}=\left(\frac{\partial\beta_k}{\partial\eta_k}-\frac{\partial\gamma_k}{\partial\eta_k}\right)x_k^ax_k^b.
\label{eq:welfare_institutional_derivative}
\eea
An institutional change is therefore locally welfare-neutral whenever
\bea
\frac{\partial\beta_k}{\partial\eta_k}=\frac{\partial\gamma_k}{\partial\eta_k}.
\label{eq:welfare_neutrality_condition}
\eea
This condition is more general than the particular zero-sum parametrization in Eq.~\eqref{eq:zero_sum_asymmetry}. It does not require $\beta_k$ and $\gamma_k$ to be equal, nor does it impose a proportional dependence on $p_k$. It only requires the marginal change in the benefit associated with privileged access to be matched by the marginal change in the cost imposed on the non-privileged player.

A useful special case is obtained when
\bea
\beta_k=\gamma_k=\eta p_k,\qquad\eta\geq0.
\label{eq:zero_sum_asymmetry_2}
\eea
In this parametrization, the asymmetric component is zero-sum at the aggregate level. Any additional benefit obtained by the player favored by the asymmetric mechanism is accompanied by an equal loss for the other player. The parameter $\eta$ determines the magnitude of this effect relative to the resource value $p_k$. The mechanism therefore redistributes the payoff associated with the shared resource without directly changing its aggregate contribution to welfare. More general choices of the dependence of $\beta_k$ and $\gamma_k$ on $\eta_k$ or on the resource characteristics can instead generate a non-zero-sum asymmetric contribution.

Although Eq.~\eqref{eq:welfare_neutrality_condition} eliminates the marginal effect of $\eta_k$ on aggregate welfare, it does not eliminate its effect on individual incentives. Holding the realized asymmetric-access state fixed, the asymmetric components of the individual payoffs respond separately to changes in $\beta_k$ and $\gamma_k$. For the player favored by the mechanism and the player who does not receive the asymmetric advantage, these effects are respectively
\bea
\frac{\partial P_{a,k}^{\mathrm{asym}}}{\partial\eta_k}=\frac{\partial\beta_k}{\partial\eta_k}x_k^ax_k^b,\qquad\frac{\partial P_{b,k}^{\mathrm{asym}}}{\partial\eta_k}=-\frac{\partial\gamma_k}{\partial\eta_k}x_k^ax_k^b.
\label{eq:strategic_institutional_derivatives}
\eea
Under the welfare-neutrality condition, these effects have equal magnitude when evaluated through the same institutional transformation, but they act in opposite directions on the two players. Consequently, the cancellation that occurs in aggregate welfare does not occur within either individual optimization problem. The institutional parameter can therefore modify the relative profitability of alternative strategies even when the welfare associated with those strategies remains unchanged.

This establishes a separation between welfare incidence and strategic incidence. Welfare incidence is determined by the net institutional effect $\beta_k-\gamma_k$, as shown by Eq.~\eqref{eq:total_welfare}, whereas strategic incidence depends separately on $\beta_k$ and $\gamma_k$ through Eq.~\eqref{eq:general_payoff_explicit}.  {
This separation is analogous to the problem of incidence in public finance: who effectively bears the effect of a rule cannot be inferred from its aggregate effect, so that the distributive and allocative consequences of an institution must be analysed separately.
} The distinction is particularly important when the institution is welfare-neutral. In such a case, changes in the access rule redistribute incentives between the players without changing the aggregate value associated with the shared resource.

The strategic consequences follow from the best-response problem defined in
Eq.~\eqref{eq:best_response}. Because its objective is the individual payoff,
changes in $\beta_k$ and $\gamma_k$ can modify the set of optimal feasible
strategies even when their net effect on Eq.~\eqref{eq:total_welfare} is zero.
In particular, increasing $\beta_k$ raises the attractiveness of selecting a
shared resource for the priority player, whereas increasing $\gamma_k$ lowers
the attractiveness of selecting that resource for the non-priority player.

As discussed in Sec.~\ref{subsec:Nash Equilibrium}, the capacity constraint makes
this effect intrinsically combinatorial, since a unilateral deviation
generally changes the interaction status of several resources simultaneously.
Consequently, the effect of an institutional parameter cannot in general be
inferred by examining a single resource in isolation. The player must compare
complete feasible knapsack allocations, and the effect of changing one
resource's payoff can depend on which alternative resources can be selected
with the remaining capacity.

This observation connects the resource-level loss condition to strategic behavior. When Eq.~\eqref{eq:general_resource_loss} is satisfied, participation in resource $k$ generates a negative direct contribution for the non-priority player. This creates a force toward avoiding that shared resource. However, the player does not necessarily abandon it because the relevant comparison is between complete feasible strategies. If the resource is retained as part of an allocation because alternative resources generate sufficiently low payoffs, its negative contribution may be offset by the rest of the bundle. Conversely, a resource whose direct contribution is positive may be rejected when selecting it prevents access to a more valuable combination of resources.

Recall that a strategy profile $(\mathbf{x}^{a*},\mathbf{x}^{b*})$ is a Nash
equilibrium if and only if it satisfies
Eqs.~\eqref{eq:nash_equilibrium}--\eqref{eq:nash_condition}, or, equivalently,
if $\Delta P_a\geq0$ and $\Delta P_b\geq0$ for every feasible unilateral
deviation, Eqs.~\eqref{eq:nash_payoff_difference_a} and
\eqref{eq:nash_payoff_difference_b}. Thus, an institutional change can affect
equilibrium behavior whenever it changes the payoff ranking between a candidate
allocation and at least one feasible unilateral deviation.
The institutional parameters enter these payoff differences through the individual incidence of shared resources. Consequently, two configurations that satisfy Eq.~\eqref{eq:welfare_equivalence} can have identical welfare values for the same strategy profile while producing different values of $\Delta P_a$ or $\Delta P_b$ for a unilateral deviation. The sign of a payoff difference can therefore change even when the corresponding change in aggregate welfare is exactly zero.

This provides the direct link between institutional asymmetry and equilibrium selection. Two institutional arrangements can generate the same welfare function over the feasible strategy space while inducing different best-response correspondences. Their Nash equilibrium sets can consequently differ even though the welfare associated with every strategy profile is identical. In particular, a welfare-neutral redistribution can make a previously unprofitable deviation attractive for one player while making the corresponding deviation less attractive for the other player.

The distinction is therefore not simply between positive and negative welfare effects. The model separates two conceptually different dimensions of institutional impact. The first is welfare incidence, determined by the net institutional contribution $\beta_k-\gamma_k$, which governs whether asymmetric access creates, preserves, or destroys aggregate value when a resource is shared. The second is strategic incidence, determined by the separate effects of $\beta_k$ and $\gamma_k$ on the individual payoff functions and therefore on the profitability of feasible deviations.

As a result, an institutional modification can leave the welfare landscape unchanged while modifying the equilibrium structure, or it can alter both welfare and strategic incentives simultaneously. The former case is especially relevant for mechanisms that redistribute access benefits and costs without changing their aggregate effect. In such cases, equilibrium differences cannot be attributed to changes in the total value generated by the allocation, but instead arise from changes in how that value and its associated costs are distributed across the players.

The model consequently distinguishes productive effects from institutional effects within the same constrained resource-allocation environment. The competitive parameter $\lambda_k$ governs the destruction of surplus generated by joint selection, the allocation coefficients $\alpha_k^i$ govern the distribution of the residual surplus, and the asymmetric-access parameters $\beta_k$ and $\gamma_k$ govern the individual incidence of the access mechanism and, depending on their net effect, its contribution to aggregate welfare. Because these channels enter the individual payoff functions differently from the aggregate welfare function, the same feasible allocation can be socially equivalent across institutional configurations while remaining strategically distinct.


\section{Social Optimum}\label{sec:Social Optimum}
The welfare function derived in Sec.~\ref{sec:Institutional Asymmetry, Welfare, and Strategic Behavior} provides the basis for evaluating the efficiency of the strategic outcomes. While the Nash equilibrium conditions identify allocations that are stable against unilateral deviations, they do not guarantee that the resulting allocation maximizes aggregate welfare. We therefore define the social optimum as the maximum welfare attainable over the feasible strategy space $\mathcal{S}=\mathcal{S}_a\times\mathcal{S}_b$:
\bea
W^\star=\max_{(x^a,x^b)\in\mathcal{S}}W(x^a,x^b).
\label{eq:social_optimum}
\eea
The set of socially optimal allocations is
\bea
\mathcal{S}^\star=\left\{(x^a,x^b)\in\mathcal{S}:W(x^a,x^b)=W^\star\right\}.
\label{eq:socially_optimal_profiles}
\eea
Since the strategy spaces are finite, at least one socially optimal allocation exists. The optimization problem in Eq. \eqref{eq:social_optimum} accounts for the capacity constraints of both players and evaluates feasible allocations according to their aggregate welfare rather than their individual profitability.

By Eq.~\eqref{eq:total_welfare}, the contribution of each resource to this objective depends on whether it is selected exclusively or jointly. Exclusive selection contributes the full resource value $p_k$, whereas joint selection contributes $(1-\lambda_k)p_k+\beta_k-\gamma_k$. The allocation coefficients
$\alpha_k^i$ do not enter the objective because they determine the
distribution of the remaining value between the players without affecting aggregate welfare.

The socially optimal allocation is therefore determined by the resource values, the competitive losses, the net effects of asymmetric access, and the capacity constraints. In contrast, the equilibrium allocation also depends on the individual distribution of payoffs, which determines the incentives to select or abandon particular resources.

To illustrate the optimization problem, we consider the asymmetric-access scenario with costs defined in Table~\ref{tab:asymmetric_cost_example}. For this parameterization, the social optimum is obtained by exhaustive enumeration of all feasible strategy profiles. Each player has a capacity of $7$, and the resource weights are $\mathbf{w}=(1,2,2,3,4)$. The resulting strategy spaces contain $22$ feasible strategies each, yielding $484$ feasible strategy profiles.

The numerical optimization gives
\bea
W^\star=18.
\label{eq:numerical_social_optimum}
\eea
This maximum is attained by twelve distinct strategy profiles, listed in Table~\ref{tab:social_optimal_allocations}. In every such profile, the two players select disjoint sets of resources. Consequently, each resource contributes its full value to aggregate welfare, and the total welfare equals the sum of the values of all available resources:
\bea
W^\star=\sum_{k=1}^{5}p_k=3+2+1+5+7=18.
\label{eq:full_value_social_optimum}
\eea
The result shows that the available capacities permit a complete allocation of the resources without overlap. In this particular example, the social optimum therefore avoids both the competitive losses associated with joint selection and the additional benefits and costs generated by asymmetric access. The institutional parameters remain relevant to individual incentives, but they do not affect the welfare attained by these socially optimal profiles.

\begin{table}[t]
\caption{Socially optimal allocations for the asymmetric-access scenario with costs defined in Table~\ref{tab:asymmetric_cost_example}. The twelve profiles attain the maximum welfare $W^\star=18$.}
\label{tab:social_optimal_allocations}
\begin{ruledtabular}
\begin{tabular}{ccc}
$\mathbf{x}^a$ & $\mathbf{x}^b$ & $W$\\
\hline
$(0,0,0,1,1)$ & $(1,1,1,0,0)$ & $18$\\
$(0,0,1,0,1)$ & $(1,1,0,1,0)$ & $18$\\
$(0,0,1,1,0)$ & $(1,1,0,0,1)$ & $18$\\
$(0,1,0,0,1)$ & $(1,0,1,1,0)$ & $18$\\
$(0,1,0,1,0)$ & $(1,0,1,0,1)$ & $18$\\
$(0,1,1,1,0)$ & $(1,0,0,0,1)$ & $18$\\
$(1,0,0,0,1)$ & $(0,1,1,1,0)$ & $18$\\
$(1,0,1,0,1)$ & $(0,1,0,1,0)$ & $18$\\
$(1,0,1,1,0)$ & $(0,1,0,0,1)$ & $18$\\
$(1,1,0,0,1)$ & $(0,0,1,1,0)$ & $18$\\
$(1,1,0,1,0)$ & $(0,0,1,0,1)$ & $18$\\
$(1,1,1,0,0)$ & $(0,0,0,1,1)$ & $18$
\end{tabular}
\end{ruledtabular}
\end{table}

The multiplicity of socially optimal profiles reflects the fact that the resources can be partitioned between the players in several ways while respecting both capacity constraints. Although these allocations generate identical aggregate welfare, they need not generate identical individual payoffs under the institutional rules of the game. Social optimality is therefore a property of the aggregate allocation, whereas its strategic stability depends on the individual incentives associated with that allocation.

We now compare the socially optimal allocations with the pure-strategy Nash equilibria obtained for the same parameterization. Table~\ref{tab:nash_welfare_efficiency} reports the four equilibria, their individual payoffs, their aggregate welfare, and the corresponding absolute welfare losses relative to the social optimum.

\begin{table}[t]
\caption{Welfare efficiency of the pure-strategy Nash equilibria for the asymmetric-access scenario with costs defined in Table~\ref{tab:asymmetric_cost_example}. The welfare loss is $\Delta W_\text{eq}=W^\star-W_\text{eq}$, with $W^\star=18$.}
\label{tab:nash_welfare_efficiency}
\begin{ruledtabular}
\begin{tabular}{cccccc}
$\mathbf{x}^a$ & $\mathbf{x}^b$ & $P_a$ & $P_b$ & $W_\text{eq}$ & $\Delta W_\text{eq}$\\
\hline
$(0,0,0,1,1)$ & $(1,1,1,0,0)$ & $12.00$ & $6.00$ & $18.0$ & $0.0$\\
$(1,0,1,1,0)$ & $(1,1,0,0,1)$ & $8.73$ & $9.07$ & $17.8$ & $0.2$\\
$(1,1,0,0,1)$ & $(0,1,1,1,0)$ & $10.62$ & $6.88$ & $17.5$ & $0.5$\\
$(1,1,0,1,0)$ & $(1,0,1,0,1)$ & $9.73$ & $8.07$ & $17.8$ & $0.2$
\end{tabular}
\end{ruledtabular}
\end{table}

The numerical results show that one of the four Nash equilibria coincides with a socially optimal allocation and attains $W^\star=18$. The remaining three equilibria generate welfare levels of $17.8$, $17.5$, and $17.8$, respectively. Their absolute welfare losses are therefore $0.2$, $0.5$, and $0.2$. In this scenario, strategic stability does not uniquely determine the socially optimal outcome: the game admits both an efficient equilibrium and equilibria with lower aggregate welfare.

The equilibrium with $W_\text{eq}=18$ involves an exclusive allocation of all five resources, as do the socially optimal profiles listed in Table~\ref{tab:social_optimal_allocations}. By contrast, the other three equilibria involve joint selection of at least one resource. Their aggregate welfare is lower because the consequences of joint selection, including competitive losses and the net effects of asymmetric access, reduce the contribution of the affected resources relative to exclusive allocation in this parameterization.

These results illustrate the distinction between the existence of a socially optimal equilibrium and the efficiency of every equilibrium. Although the institutional and competitive mechanisms permit an equilibrium that preserves the full value of the resources, they also support strategically stable allocations with positive welfare losses. The relevant question is therefore not whether the social optimum can be sustained as an equilibrium, but whether the strategic incentives ensure that all equilibrium outcomes attain it.

To examine how capacity constraints affect the relation between social optimality and strategic stability, we compare several capacity configurations while keeping all other parameters fixed at the values reported in Table~\ref{tab:asymmetric_cost_example}. In particular, the resource weights and values, the competition parameters, the allocation coefficients, the institutional benefits and costs, and the priority assigned to each resource remain unchanged. Only the capacities $C_a$ and $C_b$ vary across scenarios. This comparison allows us to distinguish the effects of aggregate capacity from those arising from its distribution between the players.

\begin{table*}
\caption{Social welfare and pure-strategy Nash equilibria under different capacity configurations. All parameters other than $C_a$ and $C_b$ are fixed at the values reported in Table~\ref{tab:asymmetric_cost_example}. The column $C_a+C_b$ gives total capacity, $W^\star$ is the maximum attainable social welfare, and $W_{\mathrm{NE}}$ lists the welfare levels of all pure-strategy Nash equilibria in descending order.}
\label{tab:capacity_comparison}
\begin{ruledtabular}
\begin{tabular}{cccccccc}
$C_a$ & $C_b$ & $C_a+C_b$ & Feasible profiles & Social optima & Nash equilibria & $W^\star$ & $W_{\mathrm{NE}}$\\
\hline
10 & 10 & 20 & 900 & 28 & 1 & 18 & $17.3$\\
10 & 7  & 17 & 660 & 20 & 1 & 18 & $17.3$\\
7  & 10 & 17 & 660 & 20 & 3 & 18 & $18,\ 17.3,\ 17.3$\\
7  & 7  & 14 & 484 & 12 & 4 & 18 & $18,\ 17.8,\ 17.8,\ 17.5$\\
7  & 5  & 12 & 308 & 4  & 4 & 18 & $18,\ 18,\ 18,\ 16.8$\\
5  & 7  & 12 & 308 & 4  & 3 & 18 & $18,\ 18,\ 16.8$\\
2  & 10 & 12 & 120 & 2  & 1 & 18 & $16.8$\\
10 & 2  & 12 & 120 & 2  & 2 & 18 & $18,\ 18$\\
2  & 7  & 9  & 88  & 1  & 1 & 15 & $15$\\
7  & 2  & 9  & 88  & 1  & 2 & 15 & $15,\ 13$
\end{tabular}
\end{ruledtabular}
\end{table*}

The total capacity $C_a+C_b$ constrains the combined weight of the resources that can be selected, but it does not by itself determine the maximum attainable welfare. The allocation of capacity between the players also matters, since the resources must be distributed between their two feasible sets while respecting both individual constraints. In the present parameterization, the total resource weight is $\sum_{k=1}^{5}w_k=12$, and the maximum welfare $W^\star=18$ is attained in every configuration with total capacity at least $12$ considered here. When total capacity falls to $9$, the maximum attainable welfare decreases to $15$. These results illustrate how capacity constraints can limit aggregate welfare even before strategic incentives are taken into account.

This feature of the model admits several institutional interpretations. In a public administration setting, the capacities may represent the limited budgets, personnel, or administrative resources available to two agencies responsible for implementing competing or complementary policies. Even when the combined resources are sufficient to support all available projects, their distribution between agencies may prevent a socially desirable allocation if some projects require a minimum level of capacity from a particular agency. A similar interpretation applies to the allocation of time and resources within households. The capacities may represent the time available to two individuals for paid work, unpaid care, and other activities, while the weights $w_k$ capture the time or effort required by each task. A sufficient total amount of time does not guarantee that all tasks can be covered if responsibilities and available time are distributed unevenly. More generally, the resources may represent educational, professional, or institutional opportunities that require different levels of effort or support. In each case, the model distinguishes between the total availability of resources and the constraints governing who can access and allocate them. These interpretations concern the feasibility of allocation; the effects of institutional priorities, unequal sharing, and penalties on individual incentives are examined separately through the payoff structure.

For each capacity configuration, we enumerate all feasible strategy profiles, determine the maximum attainable social welfare, and identify the pure-strategy Nash equilibria. Table~\ref{tab:capacity_comparison} reports the total capacity, the number of feasible profiles and socially optimal allocations, and the number and welfare levels of the pure-strategy Nash equilibria.

The results in Table~\ref{tab:capacity_comparison} reveal three distinct effects of capacity constraints on the strategic outcomes: changes in the maximum attainable welfare, changes in the multiplicity of Nash equilibria, and changes in the welfare attained at equilibrium. These effects do not follow the same pattern, which shows that the feasibility of a socially optimal allocation and its stability against unilateral deviations are distinct properties of the game.

First, the maximum attainable welfare remains $W^\star=18$ in the first eight configurations, despite substantial differences in total capacity and in its distribution between the players. In these cases, the capacity constraints permit the resources to be allocated exclusively, so that each resource contributes its full value to aggregate welfare. The result changes when total capacity falls to $C_a+C_b=9$. For both $(C_a,C_b)=(2,7)$ and $(7,2)$, the maximum welfare decreases to $15$. This reduction reflects a limitation of the feasible allocation set rather than a failure of strategic coordination: under these capacity constraints, the allocation that preserves the full value of all resources is no longer feasible. Thus, capacity constraints can reduce the social optimum itself, whereas strategic incentives determine whether the best feasible outcome is sustained in equilibrium.

Second, the results show that total capacity alone does not determine the strategic outcome. The configurations $(10,7)$ and $(7,10)$ have the same total capacity, the same number of feasible profiles, and the same number of socially optimal allocations. Nevertheless, their equilibrium structures differ. When $(C_a,C_b)=(10,7)$, the game has a unique pure-strategy Nash equilibrium with welfare $17.3$. When $(C_a,C_b)=(7,10)$, it has three equilibria, one of which attains $W^\star=18$, while the other two yield welfare $17.3$. A similar difference appears when comparing $(2,10)$ with $(10,2)$. In the first case, the unique equilibrium yields welfare $16.8$, whereas in the second, both equilibria attain the social optimum. Since all parameters other than the capacities are fixed, these differences show that the identity of the player receiving the larger capacity matters for the strategic stability of socially optimal allocations.

Third, increasing capacity does not necessarily improve equilibrium welfare or eliminate inefficient equilibria. When both capacities increase from $(7,7)$ to $(10,10)$, the number of feasible profiles rises from $484$ to $900$, and the number of socially optimal allocations increases from $12$ to $28$. However, the number of pure-strategy Nash equilibria decreases from four to one, while the lowest equilibrium welfare falls from $17.5$ to $17.3$. In particular, the unique equilibrium at $(10,10)$ remains socially inefficient despite the larger strategy spaces and the increased number of socially optimal allocations. This result illustrates that expanding the feasible strategy spaces can increase the number of allocations that attain the social optimum without improving the welfare of the equilibrium outcome. The additional strategies also modify the set of unilateral deviations available to each player, potentially destabilizing allocations that were equilibria under tighter capacity constraints.

The comparison between $(7,7)$ and the configurations $(7,5)$ and $(5,7)$ further illustrates that the relation between capacity and equilibrium multiplicity is not monotonic. With equal capacities of $7$, the game has four equilibria, with welfare levels $18$, $17.8$, $17.8$, and $17.5$. Reducing one player's capacity to $5$ does not eliminate the multiplicity of equilibria: the configuration $(7,5)$ also has four equilibria, three of which attain the social optimum, while $(5,7)$ has three equilibria, two of which attain it. Hence, a reduction in capacity can alter the number and welfare of equilibria without necessarily lowering the maximum attainable welfare.

These asymmetries must be interpreted in light of the institutional parameters in Table~\ref{tab:asymmetric_cost_example}, which remain unchanged across all capacity configurations. Player $a$ has institutional priority over resources $1$, $4$, and $5$, whereas player $b$ has priority over resources $2$ and $3$. In addition, $\alpha_{k}^a>\alpha_{k}^b$ for every resource $k$, so player $a$ receives a larger share of the residual value distributed through the allocation coefficients. However, this advantage does not imply that player $a$ always obtains a higher payoff or that allocations favoring $a$ are necessarily more efficient. The effects of joint selection also depend on the resource-specific competition losses and institutional benefits and costs. For example, resource $3$ has $\beta_3=0.2$ and $\gamma_3=1.0$, whereas resource $4$ has $\beta_4=\gamma_4=1.0$. Because these parameters remain fixed, changing the capacities changes which resources each player can select and which deviations are feasible, thereby altering how the existing institutional asymmetries shape individual incentives.

The most restrictive configurations provide a further distinction between social optimality and equilibrium stability. When $(C_a,C_b)=(2,7)$, the unique equilibrium attains the maximum welfare $W^\star=15$. When the capacities are reversed, $(C_a,C_b)=(7,2)$, the game has two equilibria, one with welfare $15$ and another with welfare $13$. The latter outcome is therefore inefficient relative to the best feasible allocation, even though the social optimum remains attainable. This comparison indicates that the distribution of capacity can determine whether the game admits only socially optimal equilibria or also supports an equilibrium with lower welfare.

Overall, the numerical results show that capacity constraints affect both the set of feasible allocations and the strategic incentives that determine equilibrium. Total capacity helps determine whether the full social value of the resources can be preserved, while the distribution of capacity between players influences the number and welfare of the resulting equilibria. Since the priority rules, allocation coefficients, competition losses, and institutional benefits and costs remain fixed, the differences across configurations arise from the interaction between capacity constraints and the model's asymmetric payoff structure. The results therefore show that neither greater total capacity nor a more balanced distribution of capacity is sufficient, on its own, to guarantee that every Nash equilibrium attains the social optimum.

\subsection{When Is Abandoning a Shared Resource Consistent with Social Optimality and Strategic Stability?}\label{subsecion:When Is Abandoning a Shared Resource Consistent with Social Optimality and Strategic Stability}

The existence of a socially optimal Nash equilibrium does not imply that all Nash equilibria attain the maximum social welfare. To examine the mechanisms underlying this distinction, we consider unilateral deviations in which one player abandons a resource selected by both players, while the other retains it and all remaining selections remain unchanged. This restricted comparison isolates the role of competition and institutional asymmetry in determining whether a shared resource is socially desirable and strategically sustainable. All deviations considered below are assumed to be feasible under the players' capacity constraints.

Throughout this subsection, we define the change in an individual payoff as the payoff after a deviation minus the payoff before the deviation,
\bea
\Delta P_i=P_i^{\mathrm{after}}-P_i^{\mathrm{before}}.
\label{eq:payoff_change_convention}
\eea
Under this convention, a player does not benefit from a deviation if and only if the corresponding payoff change is nonpositive.

Consider a resource $k$ selected by both players, with player $a$ holding institutional priority. Its contributions to the individual payoffs are
\bea
P_{a,k}^{\mathrm{sh}}&=&(1-\lambda_k)\alpha_{k}^ap_k+\beta_k,\nn\\
P_{b,k}^{\mathrm{sh}}&=&(1-\lambda_k)\alpha_{k}^bp_k-\gamma_k.
\label{eq:shared_resource_payoffs_priority_a}
\eea
If player $a$ abandons the resource, it becomes exclusive to player $b$, whose payoff contribution rises to $p_k$, while player $a$ receives zero from it. The corresponding changes in individual payoffs are
\bea
\Delta P_{a,k}^{(a)}&=&-\left[(1-\lambda_k)\alpha_{k}^ap_k+\beta_k\right],\label{eq:abandon_shared_delta_a}\\
\Delta P_{b,k}^{(a)}&=&p_k-(1-\lambda_k)\alpha_{k}^bp_k+\gamma_k.\label{eq:abandon_shared_delta_b}
\eea
Under the standard restrictions $p_k\geq 0$, $\lambda_k\in[0,1]$, $\alpha_{k}^a\geq 0$, and $\beta_k\geq 0$, the payoff change of player $a$ is nonpositive. Hence, player $a$ does not benefit from abandoning the resource when player $b$'s selection remains fixed. The payoff change of player $b$ may nevertheless be positive, reflecting the benefit player $b$ obtains when the resource becomes exclusively available. This external effect does not determine whether player $a$ has an incentive to deviate; strategic stability against unilateral abandonment must be assessed using the payoff change of the player who actually deviates.

The aggregate welfare contribution of the shared resource follows from Eq.~\eqref{eq:total_welfare} by setting $x_k^a=x_k^b=1$:
\bea
W_k^{\mathrm{sh}}=(1-\lambda_k)p_k+\beta_k-\gamma_k.
\label{eq:shared_resource_welfare_priority_a}
\eea
By contrast, exclusive allocation yields $W_k^{\mathrm{ex}}=p_k$. Defining the welfare change as the difference between exclusive and shared allocation gives
\bea
\Delta W_k&=&W_k^{\mathrm{ex}}-W_k^{\mathrm{sh}}\nn\\
&=&\lambda_kp_k-\beta_k+\gamma_k.
\label{eq:abandon_shared_delta_welfare}
\eea
Consequently, sharing yields at least as much welfare as exclusive allocation if and only if
\bea
\beta_k-\gamma_k\geq\lambda_kp_k.
\label{eq:social_optimum_shared_resource_condition}
\eea
This condition requires the net institutional contribution of joint selection to offset the competitive loss. If it is violated, exclusive allocation yields higher welfare, and any feasible transition from sharing to exclusive selection improves aggregate welfare. The condition therefore characterizes the social desirability of sharing relative to this specific alternative, rather than establishing global social optimality.

We next consider the reverse deviation, in which player $b$ abandons the resource while player $a$ retains it. The resource then becomes exclusive to player $a$, and the changes in individual payoffs are
\bea
\Delta P_{a,k}^{(b)}&=&p_k-(1-\lambda_k)\alpha_{k}^ap_k-\beta_k,\label{eq:abandon_shared_delta_a_by_b}\nn\\
\Delta P_{b,k}^{(b)}&=&-(1-\lambda_k)\alpha_{k}^bp_k+\gamma_k.\label{eq:abandon_shared_delta_b_by_b}
\eea
For player $b$ not to benefit from abandoning the resource, it is necessary and sufficient that
\bea
(1-\lambda_k)\alpha_{k}^bp_k\geq\gamma_k.
\label{eq:strategic_stability_shared_resource_condition_b}
\eea
Unlike the corresponding condition for player $a$, this inequality does not follow automatically from the standard nonnegativity restrictions. It requires the payoff retained from sharing to cover the institutional cost imposed on the non-priority player. When this condition fails, player $b$ can increase their payoff by abandoning the resource, even if sharing is socially preferable.

\begin{table}
\caption{Social efficiency and strategic stability of a shared resource.}
\label{tab:shared_resource_efficiency_stability}
\begin{ruledtabular}
\begin{tabular}{ccp{0.4\columnwidth}}
Social efficiency & Strategic stability & Interpretation\\
{\small $\gamma_k\leq\beta_k-\lambda_kp_k$} &
{\small $\gamma_k\leq(1-\lambda_k)\alpha_k^bp_k$} &
\\
\hline
Yes & Yes &
Sharing is socially preferable, and neither player benefits from abandoning the resource.\\
Yes & No &
Sharing is socially preferable, but player $b$ benefits from abandoning the resource.\\
No & Yes &
Exclusive allocation yields higher social welfare, but neither player benefits from abandoning the resource.\\
No & No &
Exclusive allocation yields higher social welfare, and player $b$ benefits from abandoning the resource.
\end{tabular}
\end{ruledtabular}
\end{table}

\begin{figure}
    \centering
    \includegraphics[scale=0.45]{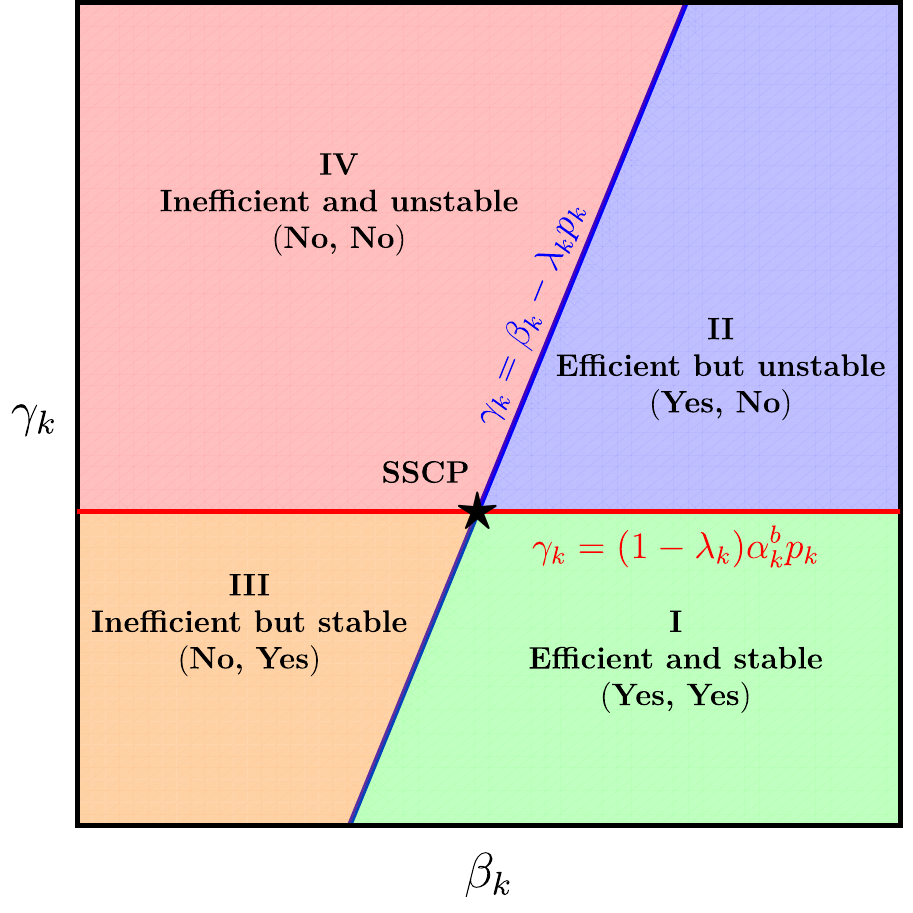}
    \caption{Social efficiency and strategic stability of a shared resource. The blue and red lines represent the thresholds in Eqs.~\eqref{eq:social_optimum_shared_resource_condition} and \eqref{eq:strategic_stability_shared_resource_condition_b}, respectively. Their intersection defines the social--strategic convergence point (SSCP). The four regions correspond to the combinations listed in Table~\ref{tab:shared_resource_efficiency_stability}.}
    \label{fig:PhaseDiagram}
\end{figure}

The joint characterization of Eqs.~\eqref{eq:social_optimum_shared_resource_condition} and~\eqref{eq:strategic_stability_shared_resource_condition_b}, summarized in Table~\ref{tab:shared_resource_efficiency_stability} and Fig.~\ref{fig:PhaseDiagram}, shows that the social desirability of sharing a resource does not necessarily coincide with the non-priority player's incentive to retain it. The first condition compares the aggregate welfare generated by shared and exclusive allocation, while the second determines whether player $b$ benefits from abandoning the shared resource. Their intersection defines the social--strategic convergence point (SSCP), at which sharing and exclusive allocation yield the same aggregate welfare and player $b$ is indifferent between retaining and abandoning the resource, with all other decisions held fixed.

The distinction between these conditions stems from the different roles of the institutional parameters. Aggregate welfare depends on the net contribution $\beta_k-\gamma_k$, whereas player $b$'s incentive to retain the resource depends on $\gamma_k$ and $\alpha_k^b$ separately. Therefore, institutional arrangements with the same effect on aggregate welfare may generate different incentives to share. The resulting divergence has a direct implication: an allocation may be socially preferable without being stable against unilateral abandonment, or remain stable even when exclusive allocation would yield higher welfare. These conditions characterize the comparison for an individual resource; establishing social optimality or a Nash equilibrium for the full game requires considering all feasible unilateral deviations under the players' capacity constraints.

\section{Summary and conclusions}\label{sec:Summary and conclusions}

This work develops a two-player knapsack game to examine how competition, institutional asymmetry, and individual capacity constraints jointly shape resource allocation, equilibrium behavior, and aggregate welfare. The analysis brings together three mechanisms that are often conflated in resource allocation problems: the destruction of surplus through competition, the distribution of the remaining surplus between agents, and the institutional rules that confer benefits on one player while imposing costs on the other. Separating these mechanisms reveals why changes in the distribution of private payoffs can have substantial strategic consequences even when their aggregate welfare effects are limited or unchanged.

A first result concerns the distinction between surplus generation and surplus appropriation. Competition over a shared resource reduces the total value generated by that resource, whereas the allocation coefficients determine how the remaining value is divided between the players. These coefficients therefore affect the distribution of private returns without directly changing aggregate welfare. Institutional benefits and costs operate differently: their net contribution determines whether asymmetric access compensates for the losses associated with competition or instead reduces total welfare. This decomposition provides an economic interpretation of the model's welfare function. It distinguishes changes in productive surplus from changes in its distribution and shows that an institutional arrangement cannot be evaluated solely by examining the private payoff of either participant.

This distinction has a direct strategic implication. Welfare equivalence does not imply strategic equivalence. Two institutional arrangements may produce the same aggregate welfare while assigning different payoffs to individual players, thereby changing their incentives to select or abandon particular resources. Since strategic choices respond to private returns rather than to aggregate welfare itself, such changes can modify best responses and the set of pure-strategy Nash equilibria without producing a corresponding change in total surplus. Institutional design thus affects not only who receives the benefits of an allocation but also which allocations can be sustained through decentralized decision-making~\cite{Wang2010TwoPerson,Wang2010TwoGroup,Wong2015,Pferschy2021}. The welfare consequences of institutional asymmetry must therefore be assessed alongside its effects on the incentives governing resource selection.

The numerical analysis further shows that these strategic effects interact with the combinatorial structure of the knapsack game. Capacity constraints do not merely limit the amount of resources that each player can select; they determine which strategic responses are feasible and how institutional asymmetries translate into equilibrium outcomes. Consequently, the distribution of capacity between players matters independently of their combined capacity. The reported configurations illustrate that reversing the distribution of capacity can change both the number of equilibria and their welfare levels, even when the underlying resource characteristics and institutional parameters remain unchanged. In particular, the configuration with capacities $(C_a,C_b)=(2,7)$ admits a unique equilibrium attaining the maximum welfare of $15$, whereas the reversed configuration $(C_a,C_b)=(7,2)$ admits two equilibria, with welfare levels of $15$ and $13$. The social optimum remains attainable in the latter case, but strategic stability also supports an inferior outcome. The relevant issue is therefore not simply whether an efficient allocation is feasible, but whether the incentives generated by the institutional and capacity structure select that allocation rather than another equilibrium.

The distinction between social efficiency and strategic stability is particularly transparent in the analysis of a shared resource. The condition in Eq.~\eqref{eq:social_optimum_shared_resource_condition} identifies when sharing generates at least as much aggregate welfare as exclusive allocation. Its economic meaning is that the net institutional contribution must be sufficient to compensate for the competitive loss associated with joint selection. This is an aggregate criterion: it evaluates the total surplus generated by the two allocation arrangements, independently of whether each player individually prefers to maintain the shared arrangement. Strategic stability, by contrast, depends on the non-priority player's private incentive to retain the resource, characterized by Eq.~\eqref{eq:strategic_stability_shared_resource_condition_b}. That condition requires the payoff obtained from sharing to cover the institutional cost borne by the disadvantaged player. Because the welfare comparison and the private incentive depend on different combinations of parameters, social desirability and strategic stability need not coincide.

The four regions in Fig.~\ref{fig:PhaseDiagram} organize this divergence into distinct economic regimes. In the region where both conditions hold, sharing is welfare-improving relative to exclusive allocation and the non-priority player has no incentive to abandon it. This is the region in which the social and private criteria are aligned for the comparison considered. Where the welfare condition holds but the stability condition fails, sharing generates greater aggregate surplus, yet the non-priority player has an incentive to abandon the resource. The allocation is socially preferable but privately unsustainable against that deviation. Conversely, when the stability condition holds but the welfare condition fails, the shared arrangement is resistant to the specified unilateral abandonment even though exclusive allocation generates greater aggregate welfare. Private incentives can therefore sustain an allocation that is socially inferior. When neither condition holds, exclusive allocation is welfare-superior and the non-priority player also benefits from abandoning the resource. The figure thus provides more than a classification of parameter values: it identifies where private incentives align with, or depart from, the aggregate welfare criterion.

The intersection of the two thresholds, referred to as the social--strategic convergence point (SSCP), marks the boundary at which the welfare difference between shared and exclusive allocation vanishes and the non-priority player is indifferent between retaining and abandoning the resource. This point clarifies that the institutional parameters governing aggregate surplus need not be the same parameters that determine an individual's willingness to participate in a shared arrangement. In particular, arrangements with equivalent aggregate welfare effects may generate different incentives because the allocation of private returns and the distribution of institutional costs remain strategically relevant. The SSCP therefore serves as a reference for comparing the welfare consequences of institutional arrangements with the incentives they create, rather than as a general guarantee of equilibrium or efficiency for the full game.

The scope of this result is important. The conditions represented in Fig.~\ref{fig:PhaseDiagram} compare shared and exclusive allocation for a given resource, with the other selections held fixed and the relevant deviation feasible under the capacity constraints. They do not, on their own, establish that a complete allocation profile is a Nash equilibrium or a global social optimum. Those conclusions require examining all feasible unilateral deviations and comparing aggregate welfare across the full set of feasible profiles. The value of the figure lies in isolating a fundamental mechanism that also matters for the larger game: the conditions that make an allocation desirable from the standpoint of aggregate welfare are not necessarily those that make it individually sustainable.

Taken together, the results establish that resource allocation under institutional asymmetry cannot be understood through resource values, capacity constraints, or payoff distribution considered in isolation. Competition determines part of the surplus available for allocation; institutional rules govern how that surplus and its associated costs are distributed; and capacity constraints determine which strategic responses are feasible. Their interaction shapes both the welfare attainable in the game and the equilibrium outcomes that decentralized behavior can support. By distinguishing welfare equivalence from strategic equivalence and by identifying the regimes in which social efficiency and strategic stability coincide or diverge, this framework provides a basis for evaluating institutional arrangements not only in terms of the aggregate outcomes they generate, but also in terms of the incentives through which those outcomes may, or may not, be sustained.

\section*{Acknowledgments} 
 JDC-Y would like to thank the Isaac Newton Institute for Mathematical Sciences (INI), Cambridge, for support and hospitality during the programme \textit{Quantum Field Theory with Boundaries, Impurities, and Defects (BID2025)}, where part of the work on this paper was undertaken. JDC-Y also acknowledges the support of the Universidad del Valle, Colombia, through the Consejo de Facultad de la Facultad de Ciencias Naturales y Exactas, which authorized his medium-term research stay at the INI.

\section*{Declaration on the Use of Artificial Intelligence}

During the preparation of this work, artificial intelligence tools, including ChatGPT and Gemini, were used as aids for writing, proofreading, and scientific cross-checking. Their use included language editing, identification of typographical errors, and consistency checks of equations, signs, notation, and mathematical expressions. These tools were not used to independently generate the scientific results or conclusions of this work. All scientific content, calculations, and interpretations were reviewed and verified by the authors, who take full responsibility for the content of the manuscript.

\bibliography{Paper/BibMochila.bib}

\end{document}